\documentclass[review]{elsarticle}

\usepackage{lineno,hyperref}
\usepackage{color}
\usepackage{graphicx}
\usepackage{subcaption}
\usepackage{caption}
\usepackage{amsmath}
\usepackage{amsfonts}
\usepackage{fullpage}
\usepackage{bm}
\usepackage{algorithm}
\usepackage{algorithmicx,algpseudocode}
\usepackage{miller}

\usepackage[normalem]{ulem}    
\usepackage{multirow}          

\usepackage{todonotes}
\biboptions{sort&compress}

\usepackage{color}

\UseRawInputEncoding 

\usepackage{dcolumn}

\journal{}

\graphicspath{{fig/},{fig/vacancy}}

\makeatletter
 \def\ps@pprintTitle{%
 \let\@oddhead\@empty
 \let\@evenhead\@empty
 \def\@oddfoot{}%
 \let\@evenfoot\@oddfoot }
\makeatother

\begin{document}

\begin{frontmatter}

\title{A multiscale and multiphysics framework to simulate radiation damage in nano-crystalline materials}

\author{Mohamed Hendy}

\author{Mauricio Ponga$^*$}
\address{Department of Mechanical Engineering, University of British Columbia, 2054 - 6250 Applied Science Lane, Vancouver, BC, V6T 1Z4, Canada}
\cortext[mycorrespondingauthor]{Corresponding author}
\ead{mponga@mech.ubc.ca}

\begin{abstract}
This work presents a multiscale and multiphysics framework to investigate the radiation-induced damage in nano-crystalline materials. The framework combines two methodologies, including molecular dynamics simulations with electronic effects and long-term atomistic diffusion simulations in nano-crystalline materials. Using this framework, we investigated nano-crystalline materials' self-healing behavior under radiation events. We found that the number of defects generated in nano-crystals during the cascade simulations was less than in single crystals. This behavior was due to the fast absorption of interstitial atoms in the grain boundary network during the cascade simulations, while vacancies migrated to the boundaries in a much longer time scale than interstitial atoms. Thus, nano-crystalline materials showed a self-healing behavior where the number and size of the defects are drastically reduced with time. 
We found that the self-healing behavior of nano-crystalline materials is limited, and about 50\% of vacancies survived. This effect resulted from clusters of vacancies' collective behavior, which are much more stable than individual vacancies.  
\end{abstract}


\end{frontmatter}

\section{Introduction}\label{Sec:Introduction}
Understanding and predicting the behavior of materials under radiation environments is paramount for a wide range of industrial applications, including next-generation of nuclear plants, especially as the world is transitioning in more environmentally friendly ways to generate energy \cite{grimes2008greater, grimes2010generating, sand2018defect, marian2017recent}. In space missions, radiation also plays an important role in determining the life of electronic components, and thus, developing materials with improving shielding capabilities is also of interest \cite{dever2005degradation}. Materials subjected to radiation environments experience a large spectrum of microstructural changes, impacting the mechanical properties of these materials over time. The processes involved in radiation-induced damage can span multiple length and time scales and are thus inherently multiscale and hierarchical \cite{Wirth:2004}. Radiation-induced damage manifests in the generation of Frenkel pair of vacancies and self-interstitial atoms (SIAs) and clusters formed from the combination of these defects. The interactions of such radiation-induced defects through migration, recombination, and annihilation dictate the degree of damage over time. The accumulation of the radiation-induced damage causes severe microstructural changes that result in mechanical degradation of the material in the form of hardening, swelling, and embrittlement \cite{ackland2010controlling}. Hence, understanding and predicting the evolution of defect clusters during radiation events is of utmost importance to design effective radiation-resistant materials. The limitations of experimental techniques require developing numerical models and simulations to understand the atomistic mechanisms of the induced radiation defects and mitigate the adverse effects of radiation-induced defects.

Due to the rapid development of successes during primary cascade events $-$that could span from a few femtoseconds (fs) to nanoseconds (ns)$-$, the \emph{in-situ} characterization of these events is highly challenging. This issue makes modeling techniques, especially molecular dynamics (MD) simulations, the preferred method to understand the fundamental aspects of radiation damage in materials.  Several MD studies have helped to understand the atomistic mechanisms during the cascade damage process in many materials, including metals and multicomponent alloys \cite{AlloysChemDis,MultiCompAlloys}, semiconductors \cite{GaN:Rad}, and others materials \cite{Zarkadoula:2020a}. MD simulations have also been used as the primary tool to understand damage mechanisms in radiation environments. Remarkably, nano-crystalline (NC) materials were studied extensively using MD simulations due to their improved radiation resistance compared to the bulk counterparts \cite{shen2007enhanced, dey2015radiation, zhang2018radiation}. In NC materials, the high misorientation angle grain boundaries (GBs) act as a trap for defects resulting from irradiation \cite{nita2004impact}. This effect has been evident in several studies where the damage due to irradiation decreased when the grain size was decreased \cite{cheng2016grain, barr2018examining}. For instance, the density of stacking fault tetrahedra (SFT) was reported much lower in the nano-crystalline materials compared to the coarser-grained material \cite{nita2005effects}. Moreover, the presence of high misorientation angle GBs results in thermal unsuitability of the radiation-induced defects in gold, palladium, and zirconium oxide \cite{CHIMI2001355, ROSE1997119}.

One possible explanation for such ability of GBs to alleviate radiation-induced defects in NC materials is that the high density of grain boundaries provides a high sink to annihilate the defects clusters. It has been shown that point-defects diffusion becomes easier in the interstitial-loaded GBs than in clean boundaries \cite{bai2010efficient, jin2018radiation}. Most MD simulations have predicted that a single GB can accommodate multiple cascades without a substantial decrease in the sink efficiency and will outperform a single crystal of the same composition with regards to radiation tolerance \cite{chen2013defect}. Another suggested mechanism is the faster migration of self interstitial atoms (SIAs) than vacancies' migration velocity towards the GBs. This faster diffusion of interstitials creates a bulk rich of vacancies and a GB network full of SIAs. Eventually, vacancies migrate in a much slower time scale compared to SIAs, and these two types of defect recombine \cite{bai2010efficient}. MD has also helped elucidate the mechanism of defects formation as it has been shown that SFT forms due to the agglomeration of the vacancy clusters via the classical Silcox and Hirsch mechanism \cite{silcox1959direct, wirth2000atomistic}.

Although classical MD has proven to be a powerful tool to model the primary cascade damage process, one drawback stems from the inability to model the electronic effects, which are essential for heat transport \cite{zarkadoula2019effects}. In materials, heat is carried out by phonons and electrons and, in metals, electrons are the primary heat carriers, representing up to 95\% of the total heat conductivity \cite{ziman_1972,makinson1938thermal}. In classical MD, phonons are explicitly resolved; hence their contribution to the system's temperature and conductivity is included. However, the electronic effects are not considered in traditional MD, limiting the applicability of MD to radiation simulations of low energy. Radiation damage problems may involve a considerable heat exchange between phonons and electrons, especially when the primary knock-on atom (PKA) energy is high \cite{nordlund1998role,lin2008electron}, deeming traditional MD results inaccurate. 

The two-temperature molecular dynamics (2T-MD) models \cite{anisimov1974electron, duffy2006including} were developed to account for the electronic temperature effects, which are absent in classical MD. Due to the importance of electronic effects during cascade damage, several 2T-MD implementations have proliferated in the recent years \cite{Tamm:2016, Tamm:2018, Tamm:2019, Eva:2019, ullah2019new, Ullah:2020, Grossi:2020, zarkadoula2018effects}. In the 2T-MD method, an electronic subsystem is modeled using a classical Fourier law to describe heat transport by the electrons. The electronic model is then coupled to the phonons in MD via a classical (e.g., non-quantum) relationship. This model can be implemented using various basis sets, including finite difference (FD) and finite element (FE) methods. In either of those approaches, the system of atoms is divided into small cells (usually called \emph{voxel}), and the Laplacian of the electronic temperature $-$required to compute heat diffusion$-$ is explicitly evaluated. Several numerical drawbacks exist with these approaches as they required the MD system to be discretized into elements to evaluate the Laplacian by recourse of a mesh. However, MD is a meshless method, and the introduction of this mesh makes the implementation seamed and laborious. At the same time, the discretization of the system in small voxels hampers the resolution of large temperature gradients during the PKA. Practically speaking, the size of these voxels has to be reduced significantly to represent the temperature gradients accurately. As a result, these voxels contain a hand full of atoms. An alternative approach preserving the true meshless nature of MD was proposed by Ullah \emph{et al.} and called $\ell$2T-MD method \cite{ullah2019new}. In $\ell$2T-MD, a Fokker-Plank equation is used to describe the energy exchange between electrons associated with atomic sites \cite{ponga2018unified}. The Fokker-Plank equation computes the probability that two near electronic sites exchange energy due to temperature differences. By taking a long-wave behavior, it can be calibrated such that  the energy  time-evolution and energy exchange match precisely the Fourier law \cite{ponga2018unified}. Thus, the use of the Fokker-Plank on top of MD allows the full atomic resolution in MD, which translates into the ability to simulate the large temperature gradients between two neighboring atoms.

Another significant disadvantage of MD is its inability to simulate long-term diffusion of defects. For instance, after forming defect clusters during the primary radiation damage, those defects interact with each other and with the GBs through thermal fluctuations and the defect's long-range elastic stress field. The defects' long-term diffusion controls the damage induced in the material and consequently is of great importance in engineering applications \cite{nordlund2019historical}. Classical MD is unable to model these long-term processes since the time scale of these events ranges from a few nanoseconds to a few seconds or more \cite{brinkman1954nature}. In order to remedy this limitation, several attempts have been made to extend the time scale accessible by MD, for example, the Hyperdynamics \cite{voter1997hyperdynamics}, Temperature Accelerated Dynamics \cite{so2000temperature}, and Parallel Replica Dynamics approach \cite{voter1998parallel} are some examples to mention but a few. While these methods are helpful for many applications, the number of atoms that can be modeled and the number of state transition events tracked down are limited. In order to overcome these challenges, several techniques have been developed, including kinetic Monte-Carlo (KMC) techniques \cite{voter2007introduction, battaile2008kinetic, Marian:KMC}, and phase-field approaches \cite{moelans2008introduction,Mianroodi:2019} which were successful in simulating the defects evolution. These methods work best for interfaces with relatively regular, ordered structures and are therefore of limited utility for non-coherent interfaces containing disordered regions, which is the case for NC materials. Furthermore, diffusive molecular dynamics (DMD) based on the Maximum entropy principle method (MXE) \cite{li2011diffusive,venturini2014atomistic,dontsova2014solute,mendez2018diffusive,mendez2020mxe} and mean-field theory has been proven to be a valuable framework to simulate the time evolution of multiple vacancies and other defects. An appealing approach of DMD is that it can be calibrated to macroscopic diffusion data of NC materials without the need to have prior knowledge of specific transition paths in these structures.

The motivation of this study is to present a (temporal) multiscale and multiphysics framework to simulate radiation damage in nano-crystalline materials. The rapid energy exchange between phonons, electrons, and PKA are described with the $\ell$2T-MD method that allows us to include electronic effects in MD. Using Ni as representative material, we investigated when the electronic effects become essential in radiation damage of materials, and compare single crystals \emph{vs.} nano-crystalline cells. To complete the description of radiation damage in nano-crystalline materials, we studied the long-term diffusion of defects using the MXE method \cite{mendez2020mxe}. We found that vacancies resting in the bulk of nanograins after the cascade simulations are absorbed by the GBs due to a combined effect of thermal excitations, differences in the chemical gradient, and long-range elastic effects generated by the defects. As a results, around $\sim50\%$ of the vacancies generated during the PKA event were annihilated at the grain boundaries, suggesting a self-healing behavior of NC materials under radiation damage. 


The paper is organized as follows. First, in Section \ref{Section:Methods}, we present the $\ell$2T-MD and MXE methodology to investigate electronic effects and the long-term behavior of materials. Next, in Section \ref{Section:Results}, we present the main results of the work. We start by comparing traditional MD with $\ell$2T-MD in bulk and NC samples. Then, we proceed to simulate the long-term behavior of the NC materials using MXE. The spatial-temporal evolution of defects is then described for time scales as long as seven days ($\sim1,800$ hrs). These results are then discussed in Section \ref{Section:Discussions}, and the manuscript is closed with the main conclusions in Section \ref{Section:Conclusions}.

\section{Methods}

\subsection{$\ell$2T-MD method} \label{l2T-MD}

Here, we describe the $\ell$2T-MD method used to include electronic effects in MD. Consider a system of $N$ atoms at finite temperature $T$. The two-temperature model assumes that the temperature field of this system can be separated into two interacting subsystems, e.g., phonons (lattice) and electrons. The lattice temperature of the $i-$th atom is denoted as $T_i^{lat}$ and its associated electronic temperature as $T_i^e$. $T_i^{lat}$ has the classical definition of $T_i^{lat}= \frac{2K}{3k_BN}$, where $K=\sum_{i=1}^N\frac{1}{2}m_i v_i^2$ is the kinetic energy of all atoms, $m_i$ and $v_i$ are the mass and the modulus of the thermal velocity of the $i-$th atom, respectively, and $k_B$ is Boltzmann's constant. Following Ullah and Ponga \cite{ullah2019new}, $T_i^e$ is assumed to fluctuate from atom to atom, and it represents the collective contribution to the electronic temperature of many electrons surrounding the $i-$th atom. 

Furthermore, $\ell$2T-MD evolves the local electronic temperature according to the following master equation \cite{ullah2019new,ponga2018unified}
\begin{equation}\label{Eq:l2t}
\dfrac{\partial T_i^e}{\partial t} =T^e_{\text{max}}\sum_{\substack{j=1\\ j\ne i}}^{N_n}K_{ij}\{\theta_j^e(1-\theta_i^e)\exp[\Delta e_{ji}]-\theta_i^e(1-\theta_j^e)\exp[\Delta e_{ij}]\} - \frac{G}{C_e}(T_i^e-T_i^{lat}) ,
\end{equation}
where $K_{ij}$ is a pair-wise exchange rate thermal coefficient for the electronic temperature between the $i-$th and $j-$th atoms whose value is defined in the sequel. $\theta_i^e = T_i^e/T^e_{\text{max}}$ is the normalized electronic temperature field for the $i-$th atom, $\Delta e_{ij}=-(\theta_i^e-\theta_j^e)$ is normalized difference in the electronic temperature between the $i-$th and $j-$th atoms. $T^e_{\text{max}}$ is an arbitrary maximum temperature value used to map $\theta_i^e ~\in ~[0,1]$. $C_e$ is the electronic heat capacity, and $G$ is the material's electron-phonon constant that defines the strength of the coupling between the lattice and electronic subsystems. $T_i^{lat}$ represents the lattice temperature computed using a cluster of atoms that surrounds the $i-$th atom. The cluster is generated using a cutoff radius that is taken equal to the cutoff of the interatomic potential for convenience. 

The pair-wise thermal exchange rate thermal coefficient can be computed using the experimentally characterized material's thermal diffusivity ($\alpha  = \kappa/(\rho_0 \cdot c_p)$, where $\kappa$ is the thermal conductivity of the material, $\rho$ its density, and $c_p$ the heat capacity at constant pressure) \cite{ullah2019new,ponga2018unified}
\begin{equation} 
K_{ij} = \frac{2 \alpha d}{Zb^2},
\end{equation}
where $d=3$ is the dimension of the problem and $Z$ is the coordination number of the pristine lattice. Since $\alpha$ depends on the temperature, so do $K_{ij}$ and $C_e$. Temperature dependence of $K_{ij}$ and $C_e$ is calculated using simple linear relation of $C_e=\lambda T_e$, and $k_e=k_0T^e/T^{lat}$ where $k_0$ is thermal conductivity at $273$ K ~\cite{hohlfeld2000electron}. At each time step $K_{ij}$ and $C_e$ are calculated using mean electronic and lattice temperature of the system.
Looking closely at Eq. \ref{Eq:l2t}, the first term on the right-hand side measures the rate of electronic energy exchanged between two adjacent particles, and it is arbitrarily selected to be around the nearest neighbors of the $i-$th particle, which makes the model \emph{local}. The second term on the right-hand side is a linear coupling term between the electrons and phonons, which calculates the amount of energy exchanged between lattice and electronic subsystems. This amount of energy has to be introduced(removed) into(from) the lattice by modifying the MD equation of motion by including a damping force $\boldsymbol F_i^{d}$. The modified equation of motion is 
\begin{equation}\label{Eq:l2t_eq_motion}
m_i \dot{\boldsymbol v}_i=\boldsymbol F_i+ \boldsymbol F_i^{d}= \boldsymbol F_i+\zeta_i m_i \boldsymbol v_i,
\end{equation}
where $\dot{\boldsymbol v}_i$ is the acceleration vector of the $i-$th atom, $\boldsymbol F_i$ is the force acting on $i-$th atom due to the interatomic interactions, and $\zeta_i m_i \boldsymbol v_i$ is the damping force that appears due to the lattice heating (cooling) coming from the electronic temperature. $\zeta_i$ is a coupling coefficient of the $i-$th atom whose strength is to be computed at each time step. The rate of work (power) done by the damping force is given by
\begin{equation}\label{Eq:l2t_eq_power}
W=\boldsymbol v_i\cdot \boldsymbol F_i^{d}= \zeta_i m_i \boldsymbol v_i\cdot \boldsymbol v_i.
\end{equation}
The work rate at the $i-$th atom has to be balanced with the amount of energy exchange by the electronic temperature per unit of time. Letting $V_{a}=\frac{V}{N}$ be the average atomic volume, the energy exchange between the lattice and the $i-$th atom is
\begin{equation}\label{Eq:l2t_eq_exchange}
\frac{\partial{e_i}}{\partial t}\biggr\rvert_{e \to lat} =GV_{a}(T_i^e-T_i^{lat}). 
\end{equation}
It is easy to see that $\zeta_i$ has to be computed at each time step to conserve energy in the system due to the energy exchange between the electrons and the phonons. The rate work done by the damping force has to be equal to the amount of energy exchange by the electrons in Eq.~\ref{Eq:l2t_eq_exchange} leading to
\begin{equation}\label{Eq:l2t_eq_zeta}
\zeta_i=\frac{GV_{a}(T_i^e-T_i^{lat})}{2K_i},
\end{equation}
where $K_i$ is the local kinetic energy of $i-$th atom. We notice that the quantity $K_i$ could be instantaneously zero at specific time steps leading to instabilities in the simulation, as the coupling coefficient will tend to infinity. This issue appears because the work done by the damping force tends to zero, and thus, even if $\zeta_i$ is large, it cannot accommodate this situation, ultimately affecting the system's stability. In order to avoid these numerical instabilities, we opted for avoiding energy exchange between phonons and electrons when $K_i< \epsilon$ at the $i-$th atom, where $\epsilon$ is a tolerance of $10^{-5}$~eV. This simple solution works well and does not affect the thermal behavior of the system. The parameters used in the $\ell2$T-MD model for Ni are specified in Table \ref{Table:material}. 

\begin{table}[H]\centering
\caption {Material properties for Ni used in the $\ell$2T-MD and MXE methods. Phonon-electron coupling constant ($G$), {\color{black}electronic heat capacity constant ($\lambda$)}, thermal conductivity at $273$ K ($k_0$), material density ($\rho_0$), and lattice heat capacity ($c_p$), formation energy ($Q_f$), migration energy ($Q_m$), attempt frequency ($\nu_0$).} \label{Table:material}
\begin{tabular}{c|c|c|c}
\hline  \hline
 Material property   & Value  & Units  & Ref. \\ \hline  \hline
 $G$    &    $3.6\times10^{17}$   &      W$\cdot$m$^{-3}$K$^{-1}$  &  ~\cite{lin2008electron} \\
 $\lambda$   &    $1065$  &      J$\cdot$m$^{-3}$K$^{-2}$ & ~\cite{lin2008electron} \\ 
  $k_0$  &   $91$  &    W$\cdot$m$^{-1}$K$^{-1}$  &  ~\cite{hohlfeld2000electron}\\ 
  $\rho_0 $   &  $8890$  &     kg$\cdot$m$^{-3}$  &  \cite{ullah2019new} \\
  $c_p $   &  $400$  &     J$\cdot$K$^{-1}$$\cdot$kg$^{-1}$  & \cite{ullah2019new}  \\
 $Q_m$   &  $1.02$  &     eV  &  ~\cite{garcia2002self} \\ 
 $Q_f$   &  $1.56$  &     eV  &  ~\cite{garcia2002self} \\ 
 $\nu_0$   &  $5.9$  &     THz$^{-1}$  &  ~\cite{debiaggi1996diffusion} \\
 
  \hline \hline
\end{tabular}
\end{table}

\subsection{Long-term diffusion} \label{Section:MXE}
After simulating the defects formed during the cascade damage, it is now required to simulate the long-term diffusion of the defects. The diffusion of defects, especially vacancies, is dominated by two factors that are material-dependent. In crystalline materials, the probability that a vacancy will migrate from one atomic site to another per unit time can be modeled using an Arrhenius law, i.e.,
\begin{equation}
\dot{\Gamma}_{i \rightarrow j} = \nu_0 \exp(-\Delta G/ k_B T),
\end{equation}
where $\nu_0$ is an attempt frequency (with units of inverse time, i.e., [time$^{-1}$]) and $\Delta G$ is the free-energy barrier that the vacancy has to overcome in order to move from one site to another. $k_B T$ is the usual thermodynamic factor. In most crystalline materials, $\Delta G \gg k_BT$, and vacancy migration is a slow process compared to the dynamics of the atoms in the material. Thus, simulation of vacancy migration is often referred as to \emph{rare event} in MD simulations beside being a common phenomenon in materials \cite{voter1997hyperdynamics,so2000temperature,voter1998parallel,voter2007introduction, battaile2008kinetic,moelans2008introduction,Mianroodi:2019}. 

In order to endeavor to simulate vacancy and interstitial diffusion in non-crystalline materials, we now briefly describe the \emph{maximum entropy} principle method (MXE)~\cite{venturini2014atomistic,mendez2020mxe} that was used to simulate the long term diffusion process. The central assumption of MXE is that the diffusion process's temporal scale is much larger than the characteristic time of the atoms' dynamics. Consequently, the microscopic state variables such as the particle's instantaneous position and its occupancy of a lattice site can be treated as random variables. The method's main idea is to supply the driving forces for state transition through a non-equilibrium statistical thermodynamics model, which is coupled with a kinetic model for the evolution of the probabilities of lattice occupancy with time. Thus, MXE bypasses the simulation of these rare events by using effective kinematic laws that can be fitted to experimental parameters such as diffusivity and migration energies \cite{mendez2020mxe}. Using this methodology, we were able to simulate the effective migration of vacancies and interstitials in NC-Ni. Now we will review the details of the method.

\subsubsection{Non-equilibrum thermodynamic model}

Consider a system of $N$ lattice sites; each can be occupied by an atom of one of $M$ species. As a consequence, the occupancy function is introduced as
\begin{equation} \label{Eq:occupancy}
n_{ik} = \begin{cases}   
1, &\ \text{if site $i$ is occupied by specie $k$ }\\
0, &\ \text{otherwise}.
\end{cases}
\end{equation}
Now, microstates of the system are described by the \emph{instantaneous} positions $\lbrace \boldsymbol q \rbrace$, \emph{instantaneous} momenta $\lbrace \boldsymbol p \rbrace$ as well as the \emph{instantaneous} occupancy $\lbrace \boldsymbol n \rbrace $. As usual in statistical mechanics, the probability distribution function is defined as a function of the three variable as $\rho(\{\mathbf{q}\},\{\mathbf{p}\},\{\mathbf{n}\})$ in the extended phase-space. This probability can be interpreted as how often is the system can be at a particular microscopic state $(\{\mathbf{q}\},\{\mathbf{p}\},\{\mathbf{n}\})$. The statistical treatment of thermodynamic systems is usually performed by invoking to phase-averaged quantities that can be determined using this probability density function (p.d.f.). Let $A(\{\mathbf{q}\},\{\mathbf{p}\},\{\mathbf{n}\})$ be an instantaneous function defined over the system. Its phase average or expected value can be computed using $\rho(\{\mathbf{q}\},\{\mathbf{p}\},\{\mathbf{n}\})$ as,
\begin{equation} \label{Eq:Entropy}
\langle A \rangle = \int_\Gamma A(\{\mathbf{q}\},\{\mathbf{p}\},\{\mathbf{n}\}) \rho(\{\mathbf{q}\},\{\mathbf{p}\},\{\mathbf{n}\}) d \Gamma.
\end{equation}
Quantities between $\langle \cdot \rangle$ represent the phase average of a given function defined over the system \cite{Landau}. In order to work with these expected values, we proceed to determine the p.d.f. of the system by using the maximum entropy principle. The next step in the MXE formalism is assuming that the system follows Jaynes' principle of maximum entropy. This principle proposes that the least biased $\rho(\{\mathbf{q}\},\{\mathbf{p}\},\{\mathbf{n}\})$ is the one that maximizes the information-theoretical entropy $S$
\begin{equation} \label{Eq:Entropy}
\max_{\rho} S = -k_B \langle \log{\rho} \rangle,
\end{equation}
which can be cast into an optimization problem to find the optimum $\rho(\{\mathbf{q}\},\{\mathbf{p}\},\{\mathbf{n}\})$. The maximization is then subjected to the two known constraints of the system, i.e., 
\begin{subequations} \label{Eq:local_constraints}
\begin{align}
& \langle {n}_{ik} \rangle = x_{ik} \\
& \langle h_i \rangle = e_i,
\end{align}
\end{subequations}
where $x_{ik}$ and $e_i$ are the $i-$th atomic molar fraction of the specie $k$ and average local energy. {\color{black}$\langle {n}_{ik} \rangle$ and $\langle h_i \rangle$ are the expected particle atomic molar fraction and expected particle energies respectively}.  $x_{ik}$ represents the probability that the $i-$th atomic site to be occupied with an atom of one of the $k$ species. When the atomic molar fraction is $x_i=1$, it represents a full certainty of the atomic site to be occupied; when $x_i=0$, it represents a full certainty of an empty site (i.e., a vacancy). $e_i$ is the local energy associated with the $i-$th site, including potential and kinetic energy. The maximization of Eq.~\ref{Eq:Entropy} is subjected to the local constraints results in a probability density function of the form
\begin{equation} \label{Eq:probability_density_function}
\rho = \frac{1}{Z} \exp^{-{ \{ \beta \}^T \{ h \} + \lbrace {\boldsymbol \gamma} \rbrace^T \{ {\bf n} \} } }
\end{equation}
with
\begin{equation} \label{Eq:partition_function}
Z = \sum_{ \{\mathbf{n}\} \in \mathcal{O}_{NM} }
 \frac{1}{h^{3N}} \int_{\Gamma} \exp^{-{\lbrace \beta \rbrace}^T {\lbrace h  \rbrace} + \lbrace {\boldsymbol \gamma} \rbrace^T \lbrace  {\bf n}  \rbrace } d\mathbf{q} d\mathbf{p},
\end{equation}
where $\rho$ and $Z$ can be regarded as the \emph{grand-canonical distribution} and \emph{grand-canonical partition function} of the system, respectively. The quantities $\{ \beta \}$ and $\{ \gamma \}$ are related to the physical quantities of atomic temperatures, $\{ T \}$, and the local particle chemical potential, $\{ \boldsymbol \mu \}$, respectively, as
\begin{equation}
T_i = \frac{1}{k_B \beta_i},
\end{equation}
and
\begin{equation} \label{Eq:ChemPot}
 {\boldsymbol \mu}_i = \frac{  {\boldsymbol \gamma}_i}{\beta_i}.
\end{equation}
It is worth mentioning that, unlike classical mechanics, in the MXE formalism, the local chemical potential and temperature are not required to be in equilibrium. This characteristic makes the method capable of modeling systems away from equilibrium. Using Eqs.~\ref{Eq:Entropy}, \ref{Eq:probability_density_function} and \ref{Eq:partition_function}, the grand-canonical total entropy is
\begin{equation} \label{Eq:PhysicalEntropy}
S =  k_B  \{ \beta \}^T   \lbrace e \rbrace - k_B \lbrace
{\boldsymbol \gamma} \rbrace^T \lbrace  {\boldsymbol x} \rbrace + k_B \log Z,
\end{equation}
and, consequently, the grand-canonical free entropy is
\begin{equation} \label{eq:Grand-Canonical-Free-Entropy}
\Phi (\lbrace \beta \rbrace, \lbrace \gamma \rbrace) = k_B \log Z (\lbrace \beta \rbrace, \lbrace \boldsymbol {\gamma} \rbrace).
\end{equation}
Equation~\ref{eq:Grand-Canonical-Free-Entropy} is very complex to solve due to the couple form of the grand-canonical partition function. Hence, a mean-field approximation is employed to derive a functional equation{\color{black}, see for instance \cite{venturini2014atomistic,mendez2020mxe}}. The mean-field grand-canonical free-entropy of the system is simplified to 
\begin{equation} \label{Eq:Grand-Canonical-Free-Energy}
\Phi_{MF} (\lbrace \beta \rbrace, \lbrace \boldsymbol{x} \rbrace) = k_B \sum_{i=1}^N \left[ \beta_i \langle h_i \rangle_0 -3 +3 \log (\hbar \beta_i \omega_i) + \sum_{k=1}^M x_{ik} \log x_{ik} \right],
\end{equation}
where $x_i$ and $\omega_i$ are the atomic molar fraction and mean-field frequency of the $i-$th site, respectively. $\langle h_i \rangle_0$ is the phase average of the interatomic potential with respect to the mean-field probability density function $\rho_0$, e.g., 
\begin{equation}
\langle h_i \rangle_0 = \dfrac{3}{2 \beta_i} +
\dfrac{|{\overline{\bf p}_i}|^2}{2m_i} + \langle V_i \rangle_0.
\end{equation}

$\langle V \rangle_0$ is the phase average of the interatomic potential, also referred to as the \emph{thermalized potential}, since atomic interactions are computed by considering the value of the positions and vibrations at a specific temperature. The non-equilibrium statistical thermodynamics model formulated in Eq.~\ref{Eq:Grand-Canonical-Free-Energy} calculates the mean positions and frequencies at a specific distribution of atomic molar fractions associated with each atomic site.

\subsubsection{Kinetic diffusion model}
The non-equilibrium statistical thermodynamics model formulated in Equation~\ref{Eq:Grand-Canonical-Free-Energy} allows for variations of the temperature and chemical potential in the system. The next step to close the system of equation is to introduce a kinetic model that computes the probabilities of lattice occupancy with time. There are many different approaches for the kinetic model, and all of them follow from phenomenological models. In this work, we follow the one used by Mendez \emph{et al.} and other works~\cite{mendez2020mxe,li2011diffusive,dontsova2014solute,mendez2018diffusive} based on a master equation for mass diffusion. The advantage of using a master equation is ability to link the phenomenological parameters in the model to well defined materials properties that can be obtained from experiments, such as vacancy diffusivity. Moreover, a versatile user-package is available with LAMMPS code, LAMMPS-MXE \cite{mendez2020mxe} which allows for further modification and development.   

Hence, the mass transport governing equation is
\begin{equation}\label{Eq:MassTransport}
\frac{\partial  x_{i}}{\partial t}  =  \sum_{{{\substack{ j=1 \\ j \neq i}}}}^{N} D_{ij}[ x_{j} (1-x_{i}) \Gamma_{i \to j}  - x_{i} (1-x_{j}) \Gamma_{j \to i} ],
\end{equation} 
where $D_{ij}$ is the pair-wise mass exchange rate that can be computed as
\begin{equation} 
D_{ij} = \frac{2d D_m}{Zb^2},
\end{equation}
where $D_m$ is the vacancy diffusion coefficient obtained from experiments. Alternatively, one can compute $D_{ij}$ as,
\begin{equation} \label{Eq:attempt_freq_mass}
D_{ij} = \nu_0 \exp (-Q_m/k_B T ),
\end{equation}
where $\nu_0$ is the {\color{black}hopping frequency}, and $Q_m$ is the migration energy, which is the energy barrier required to perform one hop of the quantity $x_i$ from the $i-$th site to the $j-$th site. In diffusion related problems, one can define a characteristic time step $\tau  = b^2/D_m \sim 1.15 \times 10^4$ s. In the MXE simulations, the integration time step used to update Eq. \ref{Eq:MassTransport} was $\Delta t = 0.2 \tau$.

The probability that to adjacent sites exchange mass is given by
\begin{equation} \label{Eq:Probability}
\Gamma_{i \to j}=\exp(\mu_{ij}/k_B T),
\end{equation} 
where $\mu_{ij}=\mu_j-\mu_i$ is the difference of the chemical potential, which is the driving force for the diffusion process. The material properties of Ni used for $\ell$2T-MD and MXE methods are summarized in {\color{black} Table}~\ref{Table:material}

After reviewing the whole model, the procedure of the implementation can be introduced. It can be divided into two main steps, a diffusive step which involves the mass transport and the structure relaxation. At first the problem is initialized by defining all the position, occupancy vectors, temperature and frequency. While fixing the atomic positions and frequencies, the atomic molar fractions are updated according to Eq.~\ref{Eq:MassTransport} in the diffusive step. It is worth mentioning that the time-step size is several {\color{black} order} of magnitude larger than the classical MD time-step. The atomic forces are calculated accordingly. Finally the positions and frequencies are updated using Eq.~\ref{Eq:Grand-Canonical-Free-Energy} by maximizing the grand-canonical free-entropy, while the atomic molar fractions are held fixed. This approach allows the model to capture multiple simultaneous transition events. 

\begin{figure}  [H]
\centering
\includegraphics[trim={0  3cm 0 3cm}, clip, width=1\textwidth]{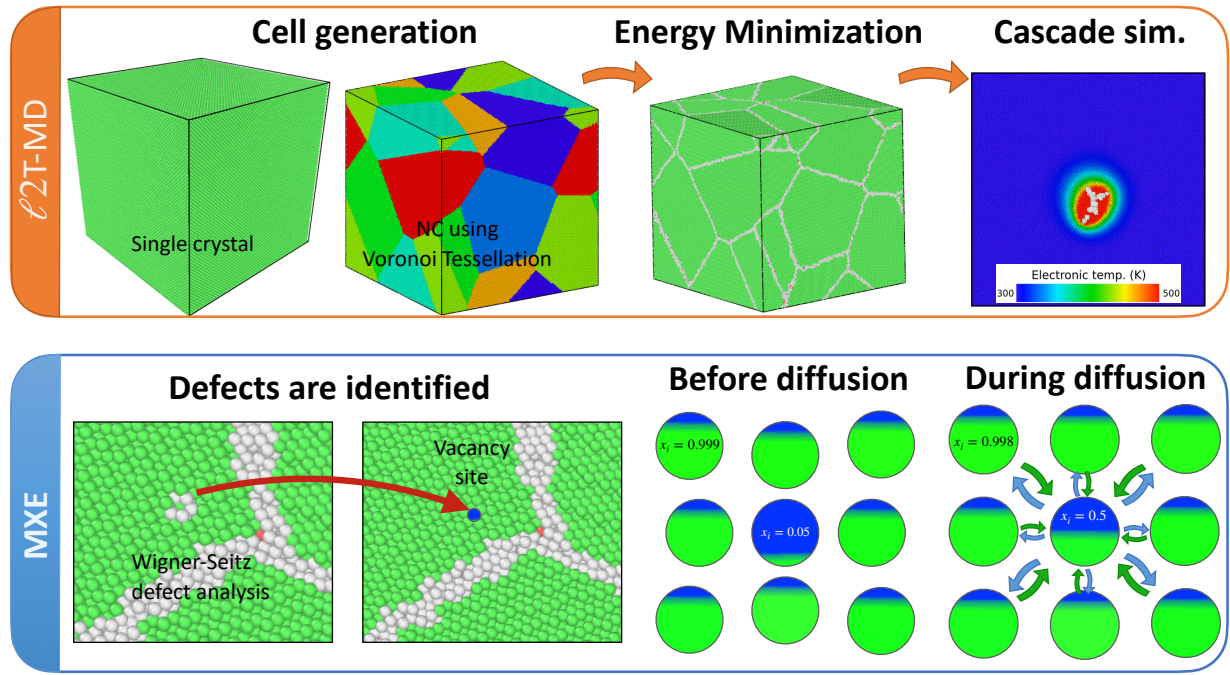}
\caption{Schematic view of the framework used in this work. In the top panel, the necessary steps to perform the $\ell$2T-MD simulations are shown, including cell generation, energy minimization, and cascade simulations. Only the last step includes the electronic temperature. In the bottom panel all necessary steps for MXE are shown.After the cascade simulations, defects are identified with the Wigner-Seitz algorithm. Vacancy sites are manually replaced by empty sites as shown in the bottom left figures. A schematic view of the atoms near the vacancy is shown in the bottom central panel, where a vacancy site is given a very low atomic molar fraction of $x_i = 0.05$.  Other atoms are given an atomic molar fraction close to one, i.e., $x_i = 0.999$. Notice that the elastic relaxation of atoms nearby the vacancy is schematically shown. When diffusion is allowed, mass is transported from the bulk to the vacancy sites, and the central atom in the bottom-right panel eventually increases its atomic molar fraction. The arrows indicate that there is an exchange of mass from/to the sites.}
\label{Fig:l2t_mxe}
\end{figure}

\subsection{Simulations set up} \label{Section:Methods}
We now describe the computational set up for the simulations, including the cascade simulations and long-term diffusive simulations carried out with $\ell$2T-MD and MXE, respectively. A schematic of the set up is shown in Fig.~\ref{Fig:l2t_mxe} for reference. MD simulations were performed using Large-scale Atomic/Molecular Massively Parallel Simulator (LAMMPS) \cite{plimpton1993fast}. The embedded atom model (EAM) potential developed by Bonny \emph{et al.} \cite{bonny2013interatomic} was used to model the inter-atomic interactions. This potential was developed to study the production and evolution of radiation defects, particularly point defects. Nickel (Ni) face-centered cubic (FCC) --with lattice parameter $a_0=0.356$ nm-- single-crystal simulation cells with side lengths {\color{black} $L=34,53$ and $71$ nm } were used as a reference {\color{black} depending on the PKA energy}. Periodic boundary conditions were employed in all three directions for all simulation cells. NC simulations cells consisted of seven randomly oriented grains generated using a Voronoi tessellation algorithm (see top left panel in Fig. \ref{Fig:l2t_mxe}). 

In order to promote the complete relaxation of the GB network in NC cells, a computational annealing procedure was used. The annealing process consisted of several cycles of high-temperature, followed by low-temperature cycles, and energy minimization procedures, similar to previous works \cite{chamani2016molecular,hasnaoui2004interaction}. Firstly, energy minimization was done at $T = 0$~K using the Polak-Ribi\`ere conjugate gradient algorithm. Then, the atoms were given an initial temperature of $T = 300$~K. After that, the simulation cell's temperature was subsequently raised from $T = 300$~K to $T =1,000$~K in $80$ picoseconds using Nos\'e$-$Hoover thermostat at zero pressure, then relaxed for $80$ picoseconds at $T = 1,000$~K then quenched to $T = 300$~K at $80$ picoseconds and relaxed for another $80$ picoseconds. The final relaxation step was subjected to another energy minimization at $T = 0$~K.

The next step was to start the cascade simulations. Cascade simulations were performed by giving one atom $-$located at the center of the cell$-$ a random high-velocity corresponding to recoil energies {\color{black} of $50$, $100$ and $150$ keV } in an NVE ensemble. This atom is often referred to as PKA. {\color{black}The electronic stopping was incorporated in the simulations through electron stopping fix in LAMMPS ~\cite{stewart2018,PhysRevB.102.024107}. The stopping-power values were calculated using the Stopping and Range of Ions in Matter (SRIM) software~\cite{ziegler2010srim} with a minimum cut-off energy of $10$~eV.} The electronic temperature effects were included using the $\ell$2T-MD method  \cite{ullah2019new}, as discussed in Section \ref{l2T-MD}. We compared the classical MD with MD plus the electronic stopping effects (referred as to MD+SRIM), and $\ell$2T-MD including SRIM (referred as to $\ell$2T-MD) simulations and assessed the effects of incorporating the electronic effects on defects evolution and final structure. For the $\ell$2T-MD simulation, the lattice and electronic subsystems were initially at $T = 300$~K before energizing the PKA. A variable time step was used, having the maximum time step restricted to $\leq 0.01$ fs for the cascade simulations to limit the maximum distance for every time-step moved by the PKA atom to $0.005$ $\AA$. At each MD step, between 10 to 15 electronic temperature integration time-steps were performed. {\color{black}The total time of the cascade simulations was about $60$ ps,
which was sufficient for the radiation-induced defects as well as the lattice and electronic temperatures to reach steady states. Each cascade simulation was repeated ten times with different random PKA directions to ensure sufficient statistical representation.}

Grain boundaries effects on the defects annihilation were assessed using the distribution of interstitial and vacancies. Defected clusters of atoms caused by the PKA were used to quantify the damage. The Ovito software was used for visualization of the defect structures \cite{ovito}, and the DXA algorithm was used for dislocation loops analysis {\color{black}with trial circuit length of $14$ and circuit stretchability value of $9$}  \cite{stukowski2012automated}.

After PKA simulations were performed with the $\ell$2T-MD method, Wigner-Seitz defect analysis was performed to determine the vacancies and SIAs using Ovito (see Fig. \ref{Fig:l2t_mxe} for a schematic view of the steps in the MXE simulations). The vacancies positions were filled with atomic sites having an atomic molar fraction of $x_i=0$ to simulate the presence of vacancies in the MXE framework, while all occupied sites $-$including interstitials$-$were assigned a value of $x_i=1$. To avoid numerical instabilities, the atomic molar fraction of the vacancies was chosen to be $x_i^v=0.05$, indicating a low probability that a Ni-atom will occupy the site. The probability of occupancy is schematically shown in Fig. \ref{Fig:l2t_mxe}, where green color in the atoms refer to highly occupied sites (i.e., $x_i\sim1$) while empty sites (i.e., $x_i\sim0$) are shown in blue. 

To exactly balance the number of vacancies in the sample, the atomic molar fraction in all occupied sites was set up to a value slightly smaller than one to compensate for this small probability of vacancies $x_i=1-\gamma$, with $\gamma = N_v/N\times0.05$, where $N_v$ is the number of vacancies identified after the PKA simulation and $N$ is the total number of atoms in the sample. For instance, using $N_a = 3,500,000$, and $N_v = 160$, $\gamma = 1.28\times 10^{-6}$, resulting in an atomic molar fraction of $x_i=0.999998$ for an occupied site.

\section{Results} \label{Section:Results} 
\subsection{Electronic temperature effects on radiation damage simulations} \label{Section:PKA}
\begin{figure}  [H]
\centering
\includegraphics[width=1\textwidth]{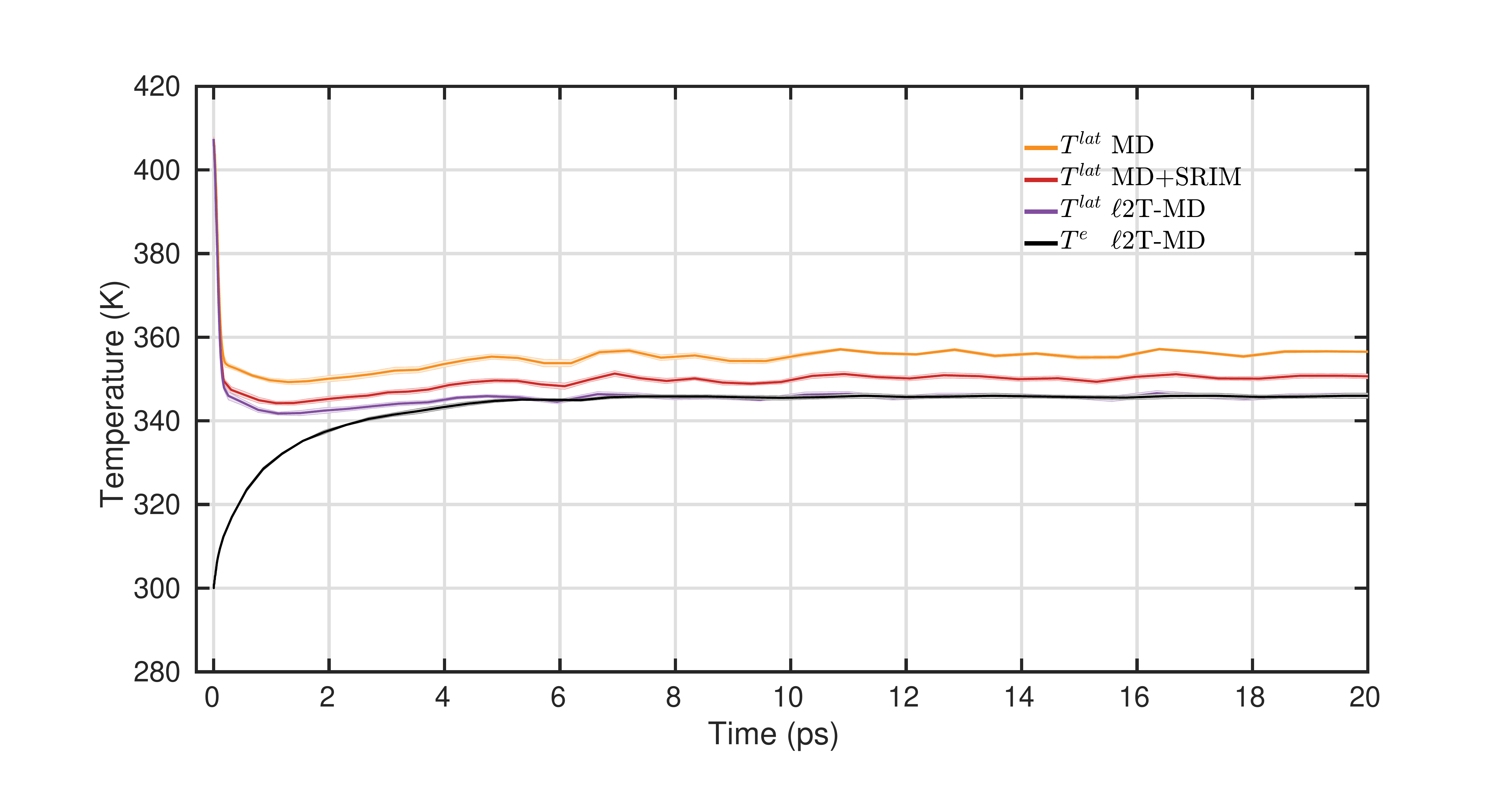}
\caption{The average lattice temperature ($T^{lat}$) evolution for MD, MD+SRIM, and for the $\ell$2T-MD simulations when a PKA with a recoil energy of {\color{black}$50$} keV was introduced in the simulations. The electronic temperature ($T^{e}$) is also shown for the $\ell$2T-MD.}
\label{Fig:lat_electronic_temp_200KeV}
\end{figure}

Before investigating the effects of the electronic subsystem on the radiation-induced defects, the lattice temperature is compared for three different systems i) classical MD $-$which will be referred to as MD$-$; ii) classical MD plus the electronic stopping effects $-$referred as MD+SRIM$-$; iii) and classical MD plus electronic stopping effects and electronic temperature effects $-$refereed as $\ell$2T-MD$-$. The temperature's time evolution for the three systems at a PKA energy of {\color{black}$50$} keV is shown in Fig.~\ref{Fig:lat_electronic_temp_200KeV}. 
Looking at the results, we observed that the lattice temperature was initially at around {\color{black}$\sim 407$} K when the PKA was inserted in the simulation and stabilized after about $\sim 5$ ps for all systems. The steady-state value of the lattice temperature was around {\color{black}$\sim356$} K for MD, followed by MD+SRIM {\color{black}$\sim350$} K, and $\ell$2T-MD {\color{black}$\sim345$} K. This behavior is expected as in the MD system, the lattice is the only mediator for heat capacity, whereas in MD+SRIM and $\ell$2T-MD electronic systems account for additional heat capacity. For the $\ell$2T-MD, we also observed that the electronic temperature increased from $300$ K to {\color{black}$\sim345$} K at around {\color{black}$\sim5$} ps and remained more or less constant after that. {\bf Supplementary Video 1} shows the time evolution of the electronic temperature for a slice of the cell for a {\color{black}$50$} keV PKA with the defect clusters superimposed in the video.  

\begin{figure} [H]
     \centering
         \centering
         \includegraphics[width=1\textwidth]{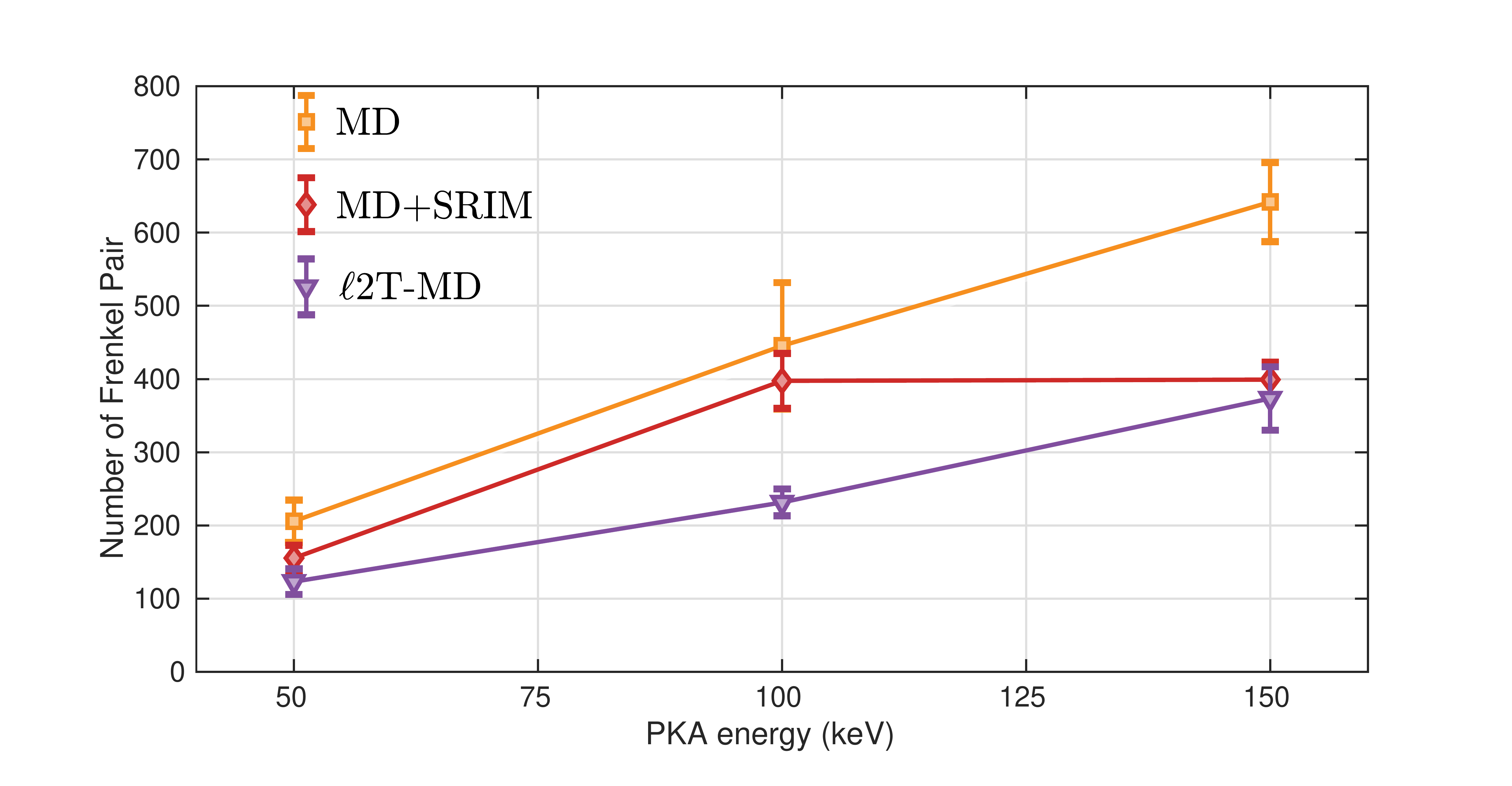}
        \caption{Frenkel pairs (FPs) at the end of cascade simulations \emph{vs.} PKA energy for MD, MD+SRIM and $\ell$2T-MD.}
        \label{Fig:steady_FP}
\end{figure}

We now focus on the number of FPs that survived the PKA responsible for the damage to the material. Figure~\ref{Fig:steady_FP} shows the number of FPs in steady-state for classical MD, MD+SRIM, and $\ell$2T-MD. At low PKA energies ($\sim 50$ keV), {\color{black}the number of FPs is significantly different between the $\ell$2T-MD and MD, while there is a small overlap between the MD+SRIM and $\ell$2T-MD}. For medium PKA energy ($\sim 100$ keV), the difference between MD and $\ell$2T-MD is more evident {\color{black},while MD+SRIM totally overlap with MD}. {\color{black}Similar behaviour was observed for the high PKA energy of $150$ keV between MD and $\ell$2T-MD, however, for MD+SRIM it overlaps with $\ell$2T-MD. For both MD and $\ell$2T-MD, the FP number increase linearly between $50$ keV and $150$ keV, however for MD+SRIM, the linear increase is broken between $100$ keV and $150$keV}. As the PKA energy increased, the number of MD predicted FPs was overestimated compared to $\ell$2T-MD. For instance, for a PKA with {\color{black}$50$} keV, MD predicts near {\color{black}$206$} FPs, whereas $\ell$2T-MD predicts {\color{black}$123$} FPs or a difference of {\color{black}$67\%$} in the surviving defects, while for MD+SRIM the overestimation of FP was around {\color{black}$26\%$} compared to $\ell$2T-MD. It is also important to remark that the FPs affect a small number of atoms in their neighborhood, changing the energy landscape. Thus, the number of atoms in defected cluster also increases significantly for MD and MD+SRIM compared to $\ell$2T-MD. 

We also point out that the MD+SRIM predicted a much closer number of FPs to {\color{black}MD} for the moderate PKA energy {\color{black}of $\sim150$ keV},{\color{black}while, for MD the overestimation of FP was around $92\%$ compared to $\ell$2T-MD} . However, as the PKA energy increased, the overestimation of MD {\color{black}compared to $\ell$2T-MD was around $72\%$}.These observations illustrate the need to include electronic effects for {\color{black}all} PKA energies studied. We also point out that the overestimation percentage is material-dependent and might differ for other materials with different heat capacity, electronic conductivity, and electron-phonon constant {\color{black}\cite{zarkadoula2014electronic, zarkadoula2015electronic, zarkadoula2017effects}}.

Generally the $\ell$2T-MD method predicts a lower amount of defects compared to classical MD even at low PKA energies of $50$ keV. The smaller number of defects is due to the energy absorbed by the electronic subsystem, which is not accounted for in classical MD. Even though the lattice temperature is not substantially different in the MD and $\ell$2T-MD systems, the energy dissipated by the electronic system resulted in a lower number of FPs than traditional MD.

\begin{figure} [H]
     \centering
     \begin{subfigure}[b]{0.3\textwidth}
         \centering
         \includegraphics[width=\textwidth, trim={7cm 0 2.8cm 0},clip]{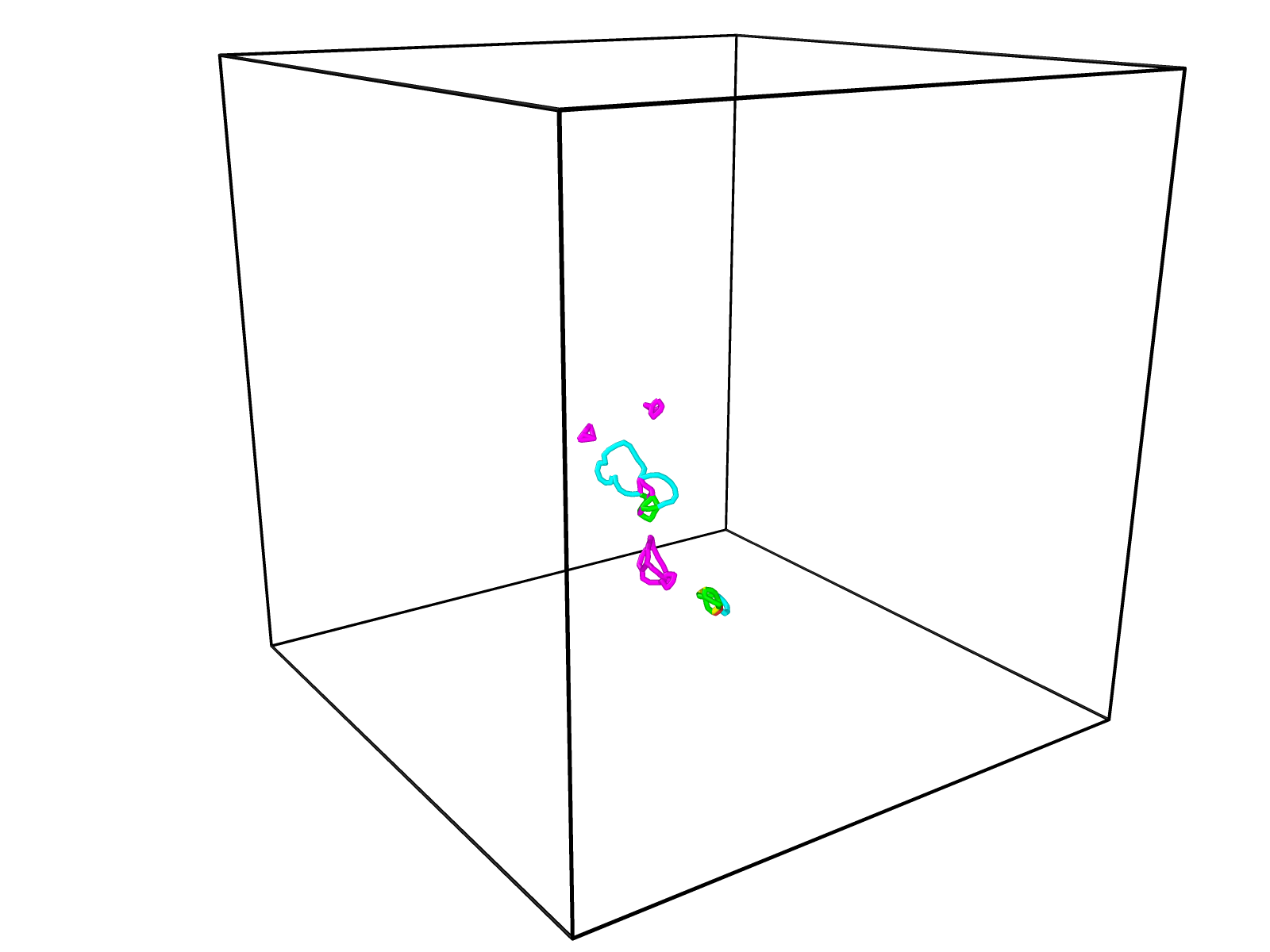}
         \caption{MD.}
         \label{Fig:dis_loop_MD_500}
     \end{subfigure}
     \hfill
     \begin{subfigure}[b]{0.3\textwidth}
         \centering
         \includegraphics[width=\textwidth, trim={7cm 0 2.8cm 0},clip]{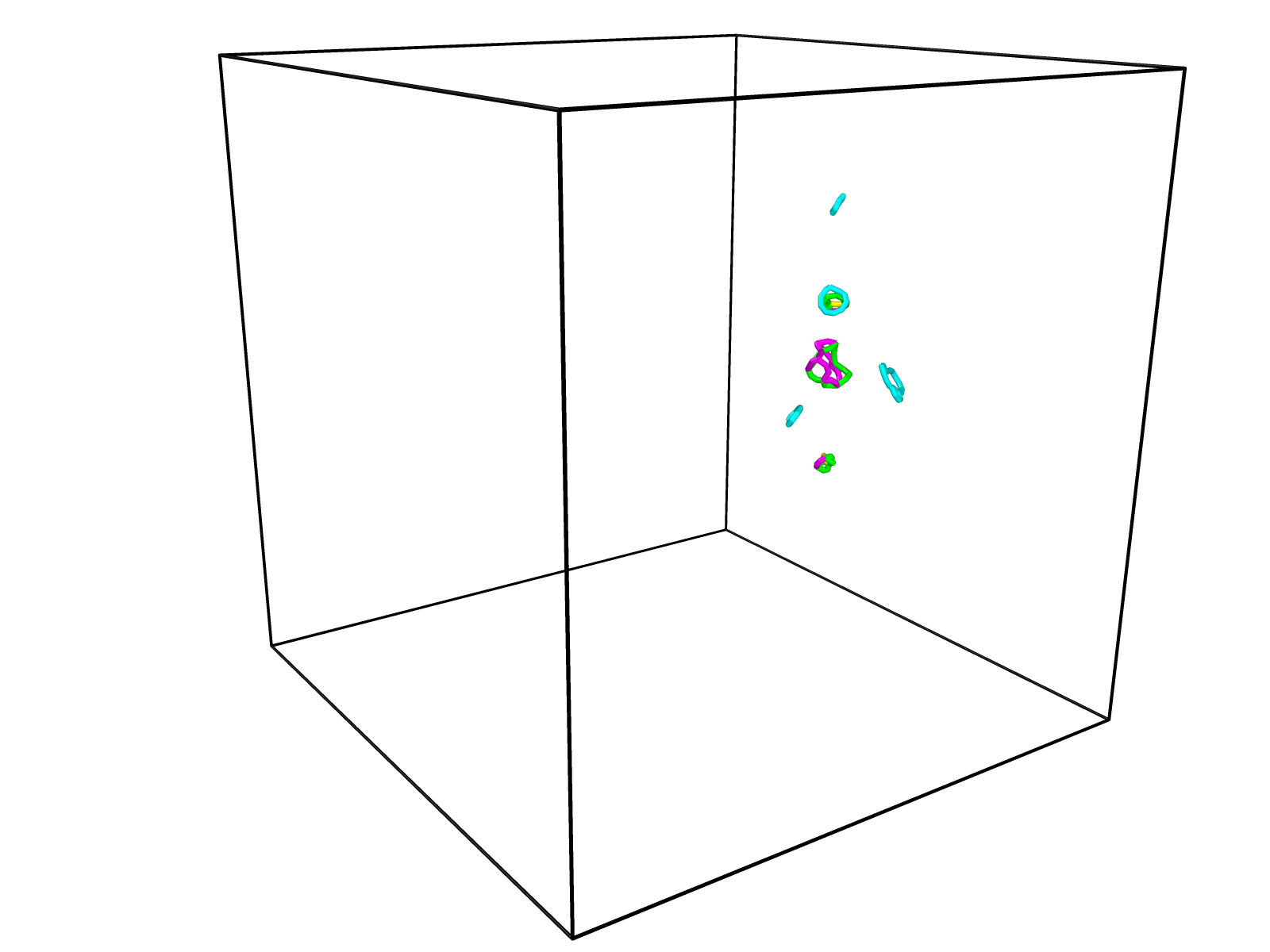}
         \caption{MD+SRIM.}
         \label{Fig:dis_loop_SRIM_500}
     \end{subfigure}
     \hfill
     \begin{subfigure}[b]{0.3\textwidth}
         \centering
         \includegraphics[width=\textwidth, trim={7cm 0 2.8cm 0},clip]{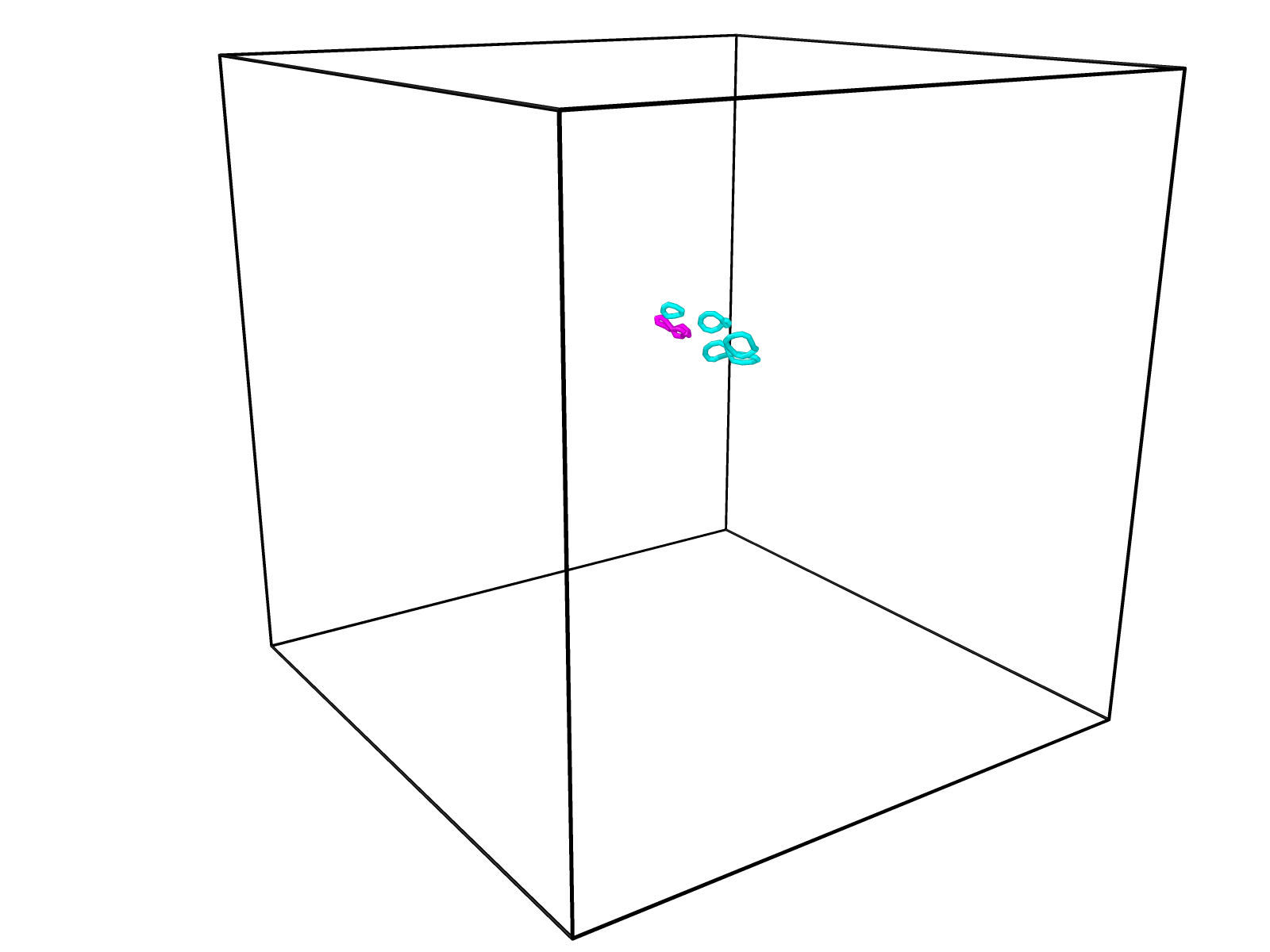}
         \caption{$\ell$2T-MD.}
         \label{Fig:dis_loop_l2t_500}
     \end{subfigure}
     \hfill
     \begin{subfigure}[b]{0.3\textwidth}
         \centering
         \includegraphics[width=\textwidth, trim={7cm 0 2.8cm 0},clip]{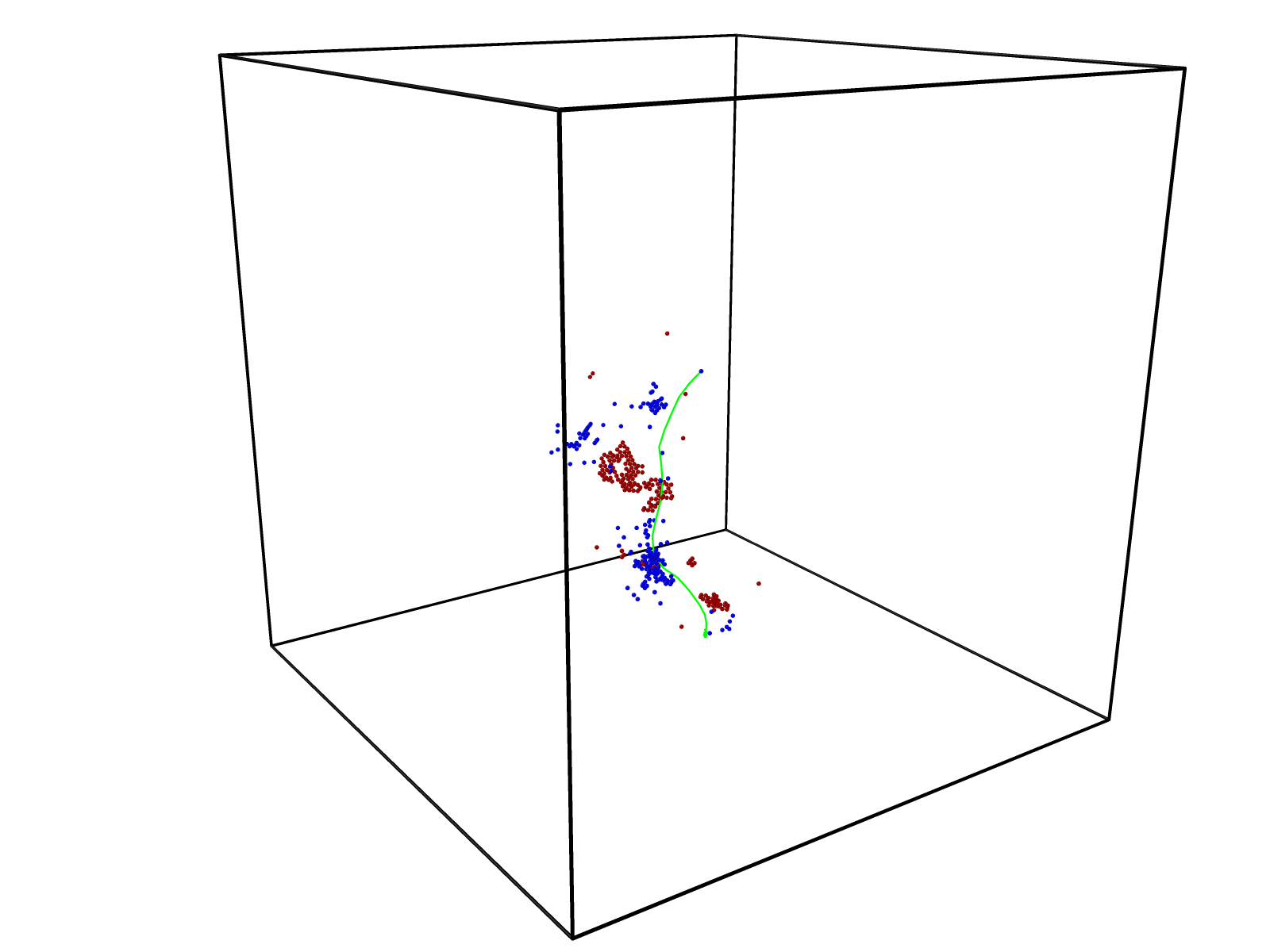}
         \caption{MD.}
         \label{Fig:clusters_MD_500}
     \end{subfigure}
     \hfill
     \begin{subfigure}[b]{0.3\textwidth}
         \centering
         \includegraphics[width=\textwidth, trim={7cm 0 2.8cm 0},clip]{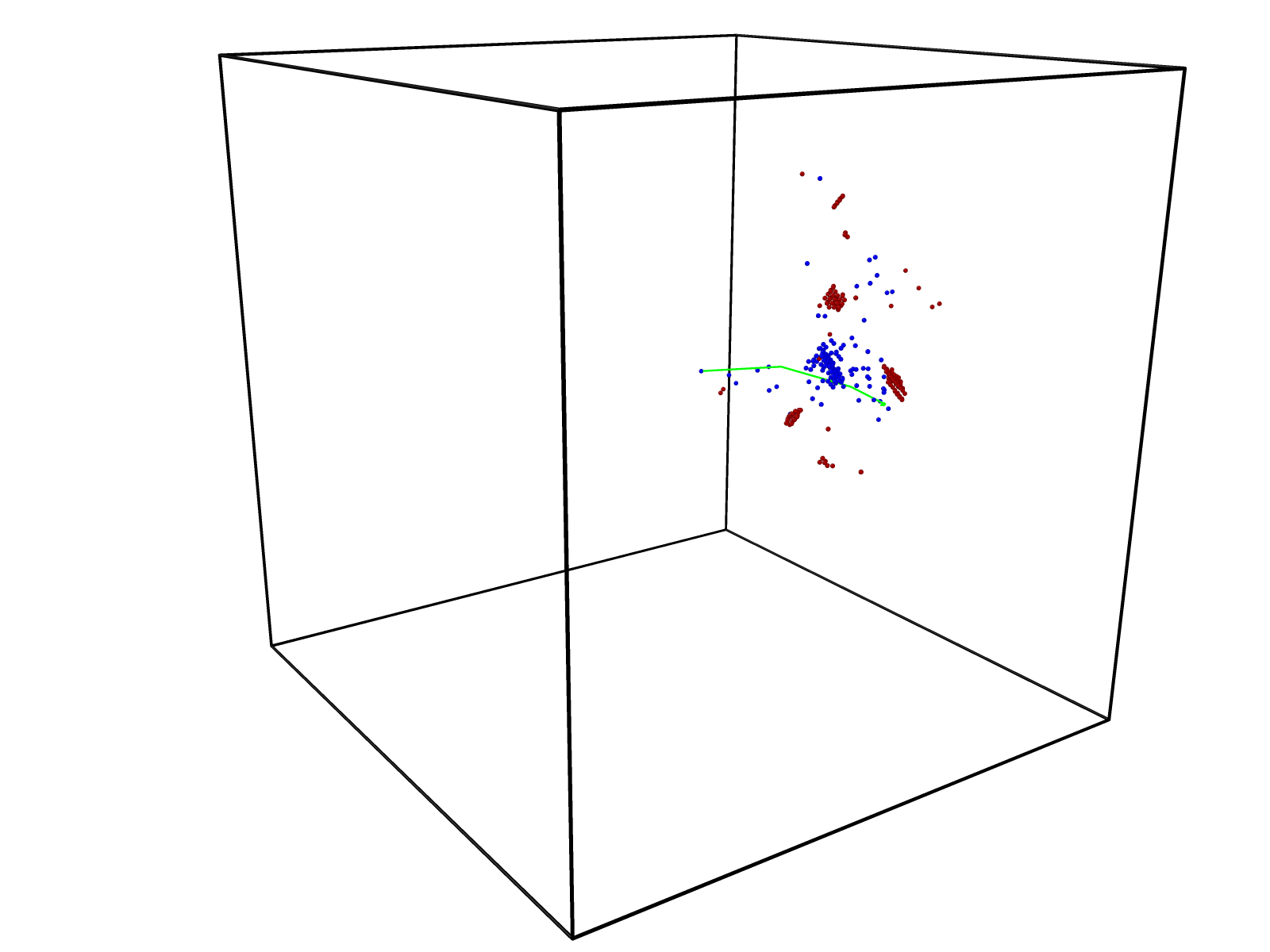}
         \caption{MD+SRIM.}
         \label{Fig:clusters_SRIM_500}
     \end{subfigure}
     \hfill
     \begin{subfigure}[b]{0.3\textwidth}
         \centering
         \includegraphics[width=\textwidth, trim={7cm 0 2.8cm 0},clip]{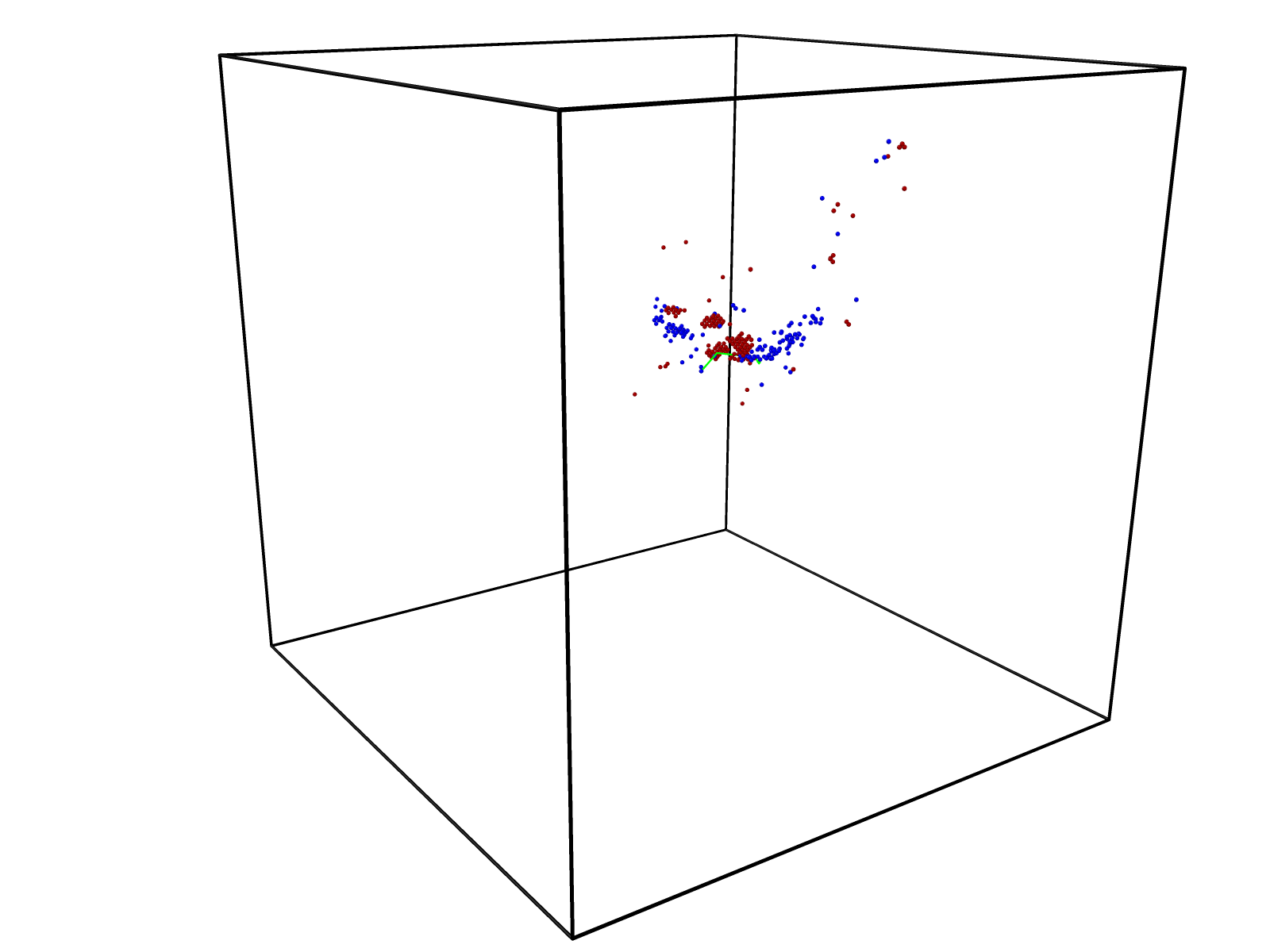}
         \caption{$\ell$2T-MD.}
         \label{Fig:clusters_l2t_500}
     \end{subfigure}
        \caption{Dislocation loops and {\color{black}FPs} after the cascade simulations using a 50 keV PKA for MD, MD+SRIM, and $\ell$2T-MD. Top figures (a)-(c) show the distribution of dislocation loops in the simulation cell. Dislocation loops are colored according to their Burgers vectors, where green denotes a Shockley partial dislocation, purple a stair-rod dislocation, light-blue a Frank dislocation, yellow a Hirth dislocation, and dark-blue a perfect dislocation. Bottom figures (d)-(f) show the distribution of {\color{black}vacancies (blue) and SIAs (red)} in the simulation cell, along with the PKA path in green.}
\label{Fig:srim_l2t_SIA_vac_dis_single_500keV}
\end{figure}

Having quantitatively compared the number of FPs between the {\color{black}three} techniques, we now focus our attention on the distribution and clustering of the defects. Figure~\ref{Fig:srim_l2t_SIA_vac_dis_single_500keV} shows only the spatial distribution of the dislocation loops originated from the cascade simulation, along with the SIAs and vacancies for MD, MD+SRIM, and $\ell$2T-MD {\color{black}for a recoil energy of $50$ keV}. Figures ~\ref{Fig:srim_l2t_SIA_vac_dis_single_500keV}(d)-(f) shows the {\color{black}corresponding vacancies (blue) and SIAs (red), which some clusters and forms the  dislocation loops}  after the cascade simulation.  It was observed that the single crystal contained defects in the form of dislocation loops, as shown in Fig.~\ref{Fig:srim_l2t_SIA_vac_dis_single_500keV}, and individual interstitial and vacancies accompanied these loops. When the electronic effects were taken into account, they altered the dislocation loops size and configuration compared to MD (cf. Fig. \ref{Fig:srim_l2t_SIA_vac_dis_single_500keV}(a) with Fig. \ref{Fig:srim_l2t_SIA_vac_dis_single_500keV}(b) and Fig. \ref{Fig:srim_l2t_SIA_vac_dis_single_500keV}(c) for dislocation loops in the MD, MD+SRIM, and $\ell$2T-MD simulations, respectively). In the classical MD simulations, the defects clustered {\color{black}along} the trajectory of the PKA and generated much bigger dislocation loops (see Fig. \ref{Fig:srim_l2t_SIA_vac_dis_single_500keV}(a)). On the other hand, for MD+SRIM and $\ell$2T-MD, dislocation loops were smaller in size and aligned with the PKA path, {\color{black}while the FPs were more scattered away from the PKA path rather than concentrated}  as shown in Fig. \ref{Fig:srim_l2t_SIA_vac_dis_single_500keV}(b) and Fig. \ref{Fig:srim_l2t_SIA_vac_dis_single_500keV}(c), respectively.  

\begin{figure} [H]
\centering
\includegraphics[width=1\textwidth]{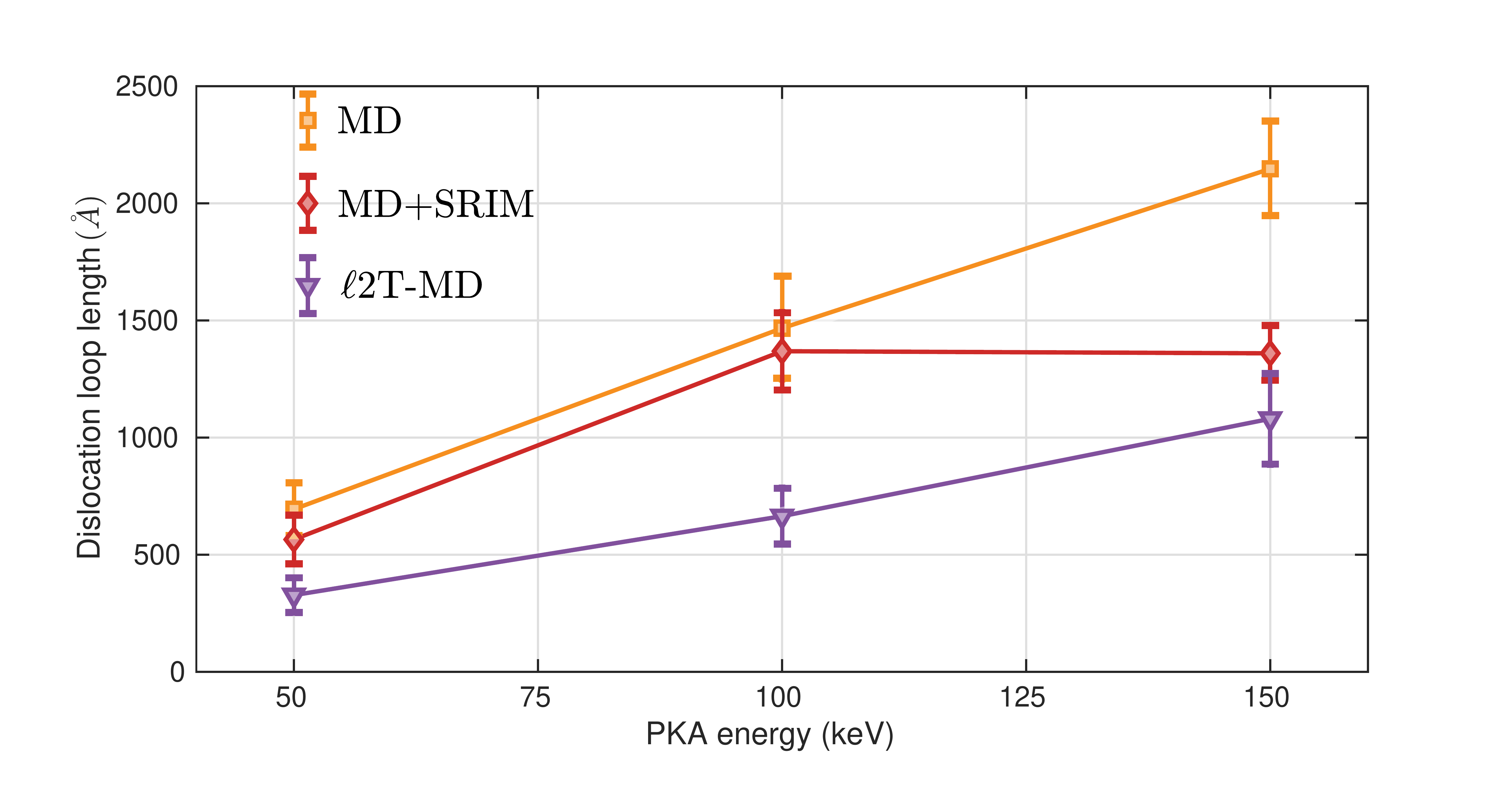}
\caption{Dislocation loop length resulted from the radiation-induced defects \emph{vs.} PKA energy for MD, MD+SRIM and $\ell$2T-MD.}
\label{Fig:dislocation_length_single}
\end{figure}

Figure~\ref{Fig:dislocation_length_single} shows the dislocation {\color{black}loop length} for different PKA energies and methods. The dislocation {\color{black}loop length} correlates well with the evolution of the FP shown in Fig. \ref{Fig:steady_FP}. MD overestimates the $\ell$2T-MD {\color{black}for all PKA energies}, while MD+SRIM {\color{black}behaviour depended on the PKA energy}. Generally, MD simulations showed more defects clusters than the $\ell$2T-MD simulations. For instance, at a recoil energy of {\color{black}$50$} keV, MD predicted a dislocation {\color{black}length} of {\color{black}$694$ \AA}, while $\ell$2T-MD predicted {\color{black}$327$ \AA}. Besides, the clusters generated in the $\ell$2T-MD simulations were smaller in size, indicating that the classical MD overestimates the size and shape of the defect clusters formed and their number. Moreover, the distribution of defects clusters is different in $\ell$2T-MD compared to classical MD, similar to {\color{black}\cite{zarkadoula2017two, zarkadoula2016effects}}.

\subsection{Primary radiation damage in nanocrystalline Ni} \label{Section:PKA-NC}

After studying the effects of radiation in single Ni crystals using classical MD, MD+SRIM and $\ell$2T-MD and showing that electronic effects are important when the PKA energy is high, we now endeavour to investigate the effect of radiation in the damage of NC materials. To do that, we systematically compared the results between a Ni-single crystal and a Ni-NC with $\ell$2T-MD at {\color{black}$50$} keV. NC cells were produced using the procedure described in Section \ref{Section:Methods}. 
 
\begin{figure} [H]
     \centering
     \begin{subfigure}[b]{0.48\textwidth}
         \centering
         \includegraphics[width=\textwidth]{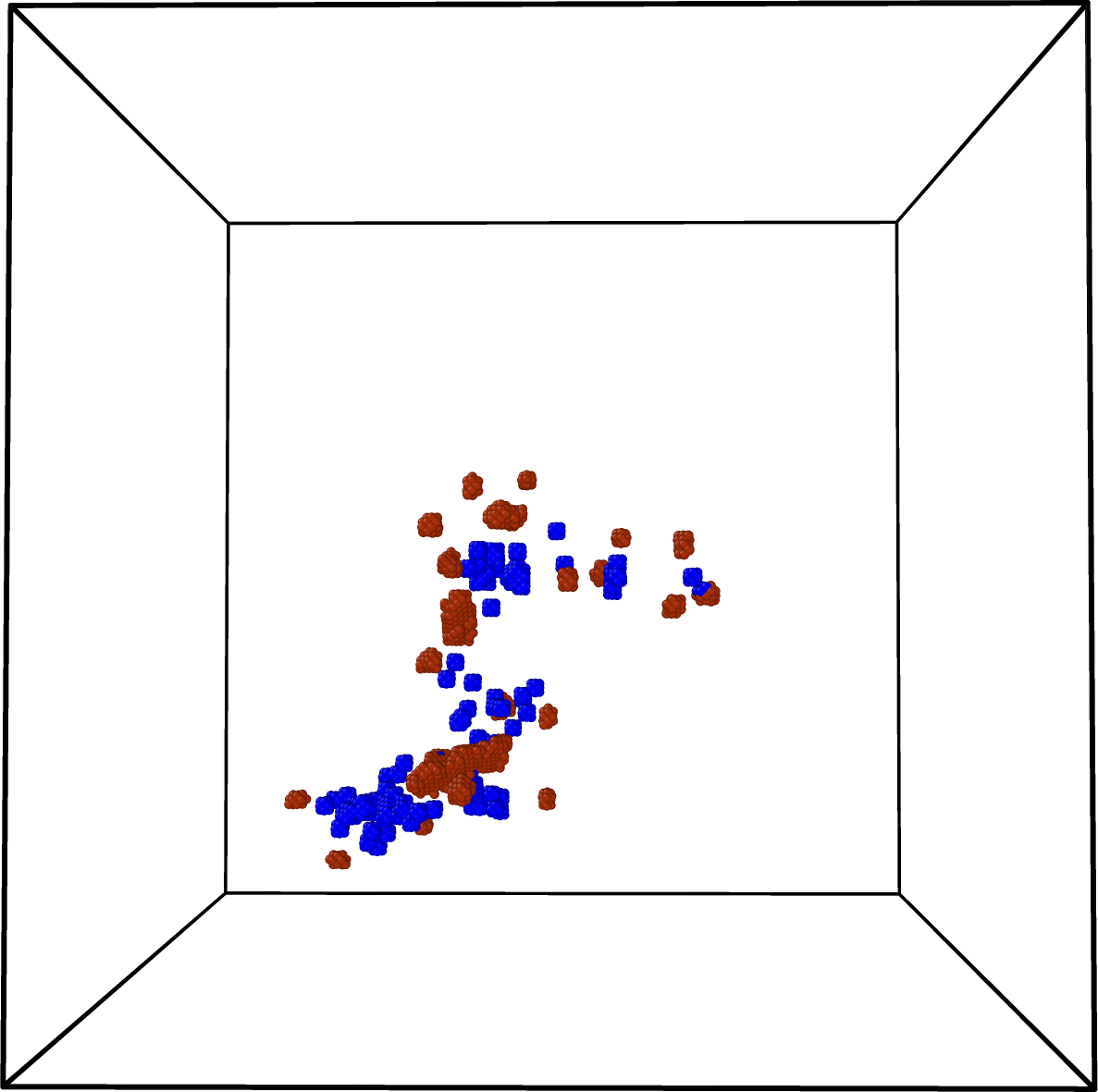}
         \caption{Single-crystal cell.}
         \label{Fig:single_clusters_l2t_500keV}
     \end{subfigure}
     \hfill
     \begin{subfigure}[b]{0.48\textwidth}
         \centering
         \includegraphics[width=\textwidth]{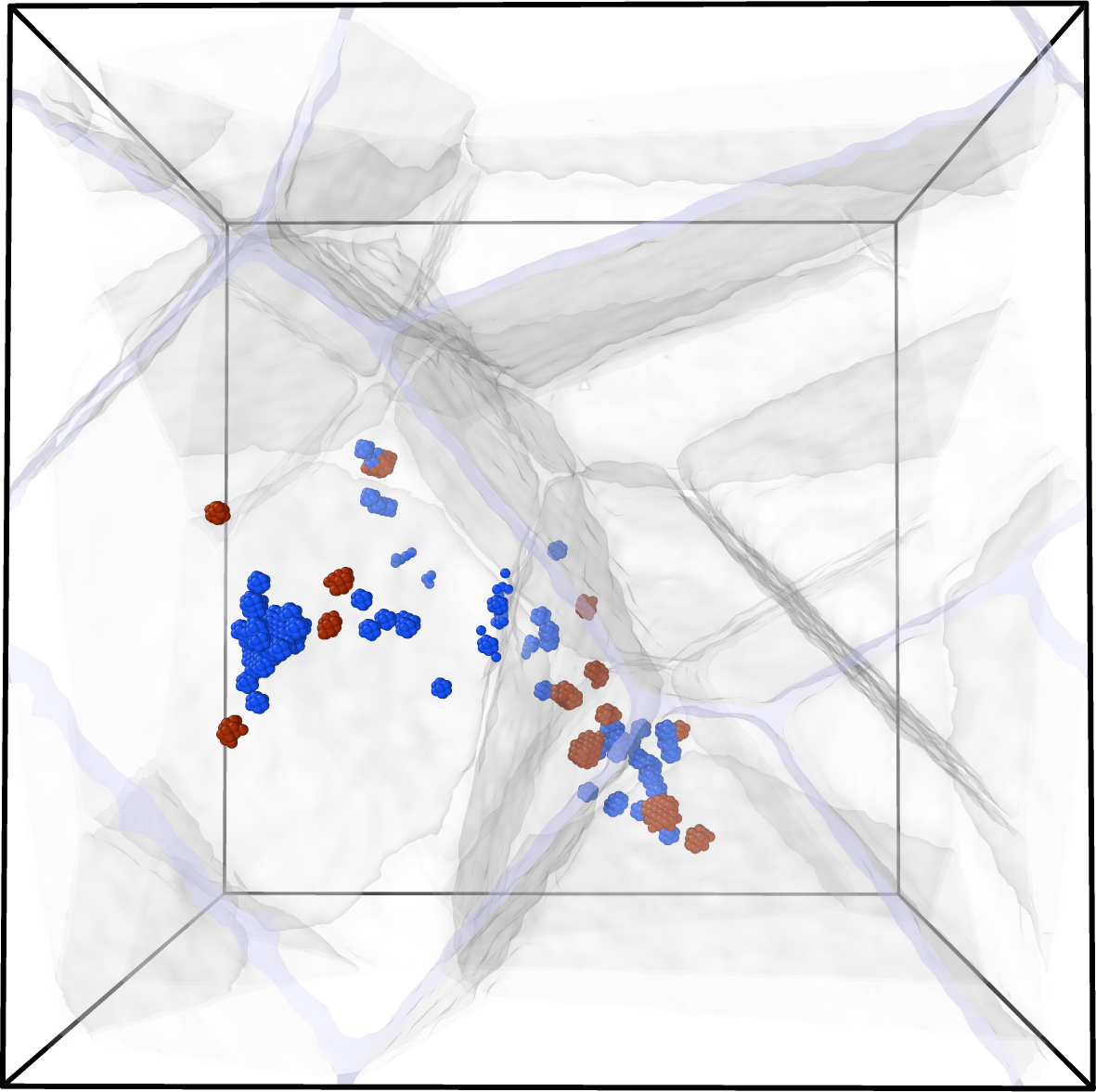}
         \caption{Nano-crystalline cell.}
         \label{Fig:7G_clusters_l2t_500keV}
     \end{subfigure}
        \caption{Defect clusters after a cascade simulation with a $50$ keV PKA for (a) a single crystal and (b) nano-crystalline cell using the $\ell$2T-MD method. The grain boundary network of the nano-crystalline sample is shown as a transparent surface network.}
        \label{Fig:clusters_NC_single_500keV}
\end{figure}

Figure~\ref{Fig:clusters_NC_single_500keV} shows the computational cells for a single crystal (Fig.~\ref{Fig:clusters_NC_single_500keV}(a)) and a NC containing 7 grains (Fig.~\ref{Fig:clusters_NC_single_500keV}(b)) that have been damaged with a PKA with a recoil energy of {\color{black}$50$} keV. In the figure, all atoms have been removed except the ones in defect clusters generated after the PKA {\color{black}coloured according to its type, blue for vacancy and red for SIA}, and the GBs in Fig.~\ref{Fig:clusters_NC_single_500keV}b were plotted using a transparent mesh. The GBs result in a complex network in the computational cell. Next, the PKA produced FPs, which in turn can coalesce and generate larger defect clusters {\color{black}of defective atoms}. {\color{black}It is worth mentioning that the generated FPs cause the distortion of the lattice around them leading to the formation of non-FCC atoms around them, referred here as defective atoms}. Even though the cell size and PKA energy were the same for both cases, the number, shape, and size of the {\color{black}SIA defect clusters} changed drastically in the simulations as portrayed in Fig.~\ref{Fig:clusters_NC_single_500keV} (cf. (Fig.~\ref{Fig:clusters_NC_single_500keV}(a) and Fig.~\ref{Fig:clusters_NC_single_500keV}(b)). A visual inspection of the {\color{black}SIA} defects after the PKA revealed that the defect clusters had smaller sizes for the NC case. {\color{black}The largest defect cluster for SC was formed of $272$ atoms, while for NC was $84$, indicating a significant effect of the GB presence on the size of the SIA defect clusters. Furthermore, the total number of defective atoms around SIAs for the NC is $1062$ while it is $406$ for NC. Looking at the number of the defect clusters, it was $23$ for SC and 13 for NC as shown in table \ref{tab:Defect_table}. On the other hand, the vacancy defect clusters showed the similar behaviour for SC and NC. For instance, the number of defective atoms is $1162$ and $1069$ respectively with overlapping standard error values as shown in table \ref{tab:Defect_table}. Similar trend is observed for the largest defect cluster with $402$ for SC and $382$ for NC. For the number of defect vacancy clusters, there was a slight decrease in NC compared to SC from $33$ to $29$. 

This suggests that the GBs acts as a quick sink for the SIAs in just after the cascade event, while it doesn't offer much improvement during the same period of time for the vacancy defects. This behaviour suggests that SIAs can be quickly reabsorbed in the GBs during the PKA event, while vacancies have a long-living time. Since SIAs are absorbed in the GBs, there are fewer interstitial atoms in the bulk of the grains, ultimately impacting the size of large SIA clusters. As a result, the number and size of the dislocation loops observed in NC samples are much smaller than SC. Interestingly, the size and number of clusters are always smaller in NC samples. The collective behavior of SIAs illustrates the beneficial effect of the GBs compared to SC cells during the PKA simulations. However, it does not explain the reduction in vacancy defects. {\bf Supplementary Video 2} shows the time evolution of the electronic temperature for a slice of a nano-crytalline cell with the defect clusters superimposed in the video.} 

\begin{figure} [H]
     \centering
     \begin{subfigure}[b]{0.98\textwidth}
         \centering
     \includegraphics[width=1\textwidth]{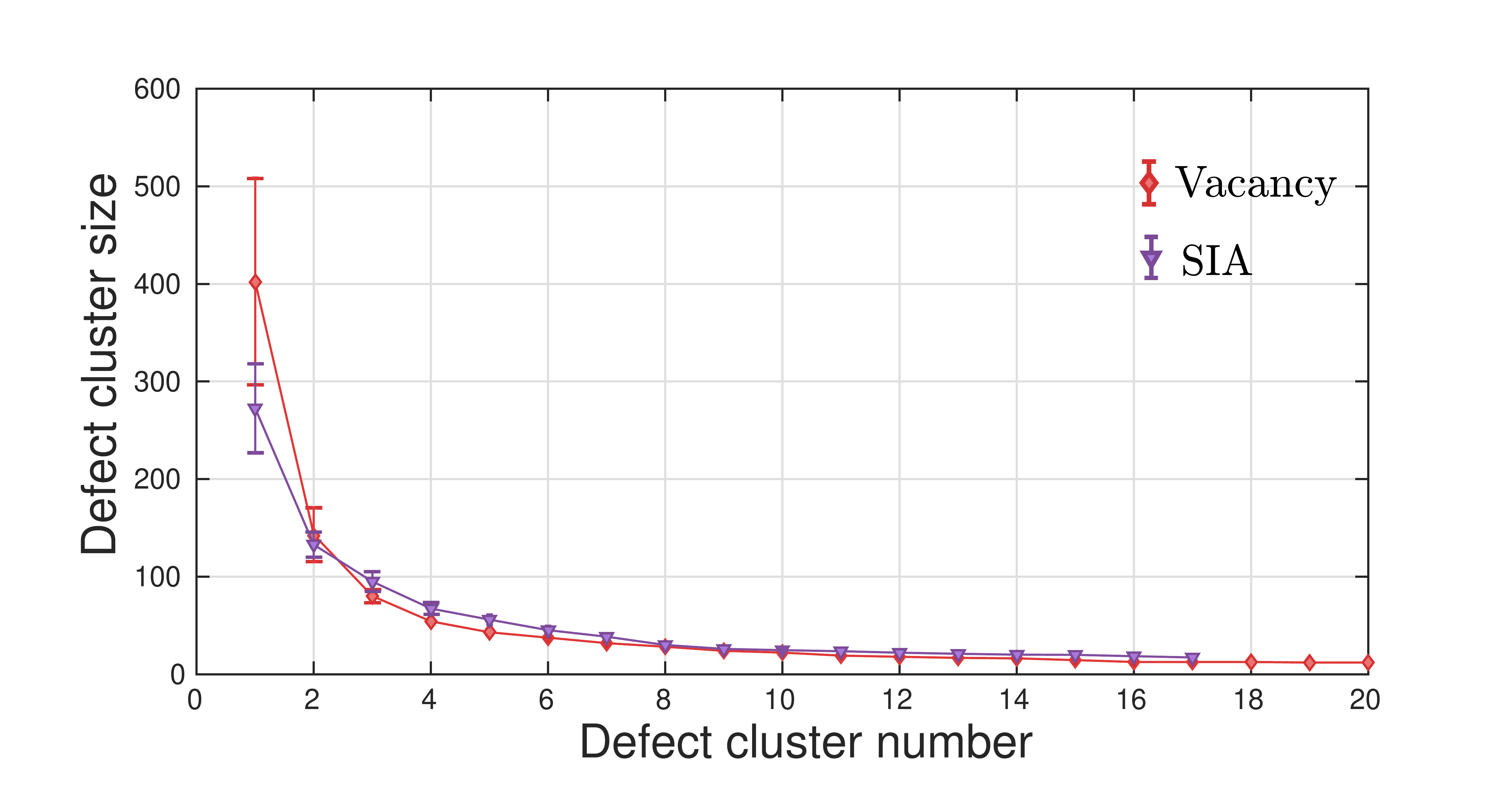}
         \caption{Single-crystal cell.}
          \label{Fig:cluster_count_single}
     \end{subfigure}
     \hfill
     \begin{subfigure}[b]{0.98\textwidth}
         \centering
         \includegraphics[width=\textwidth]{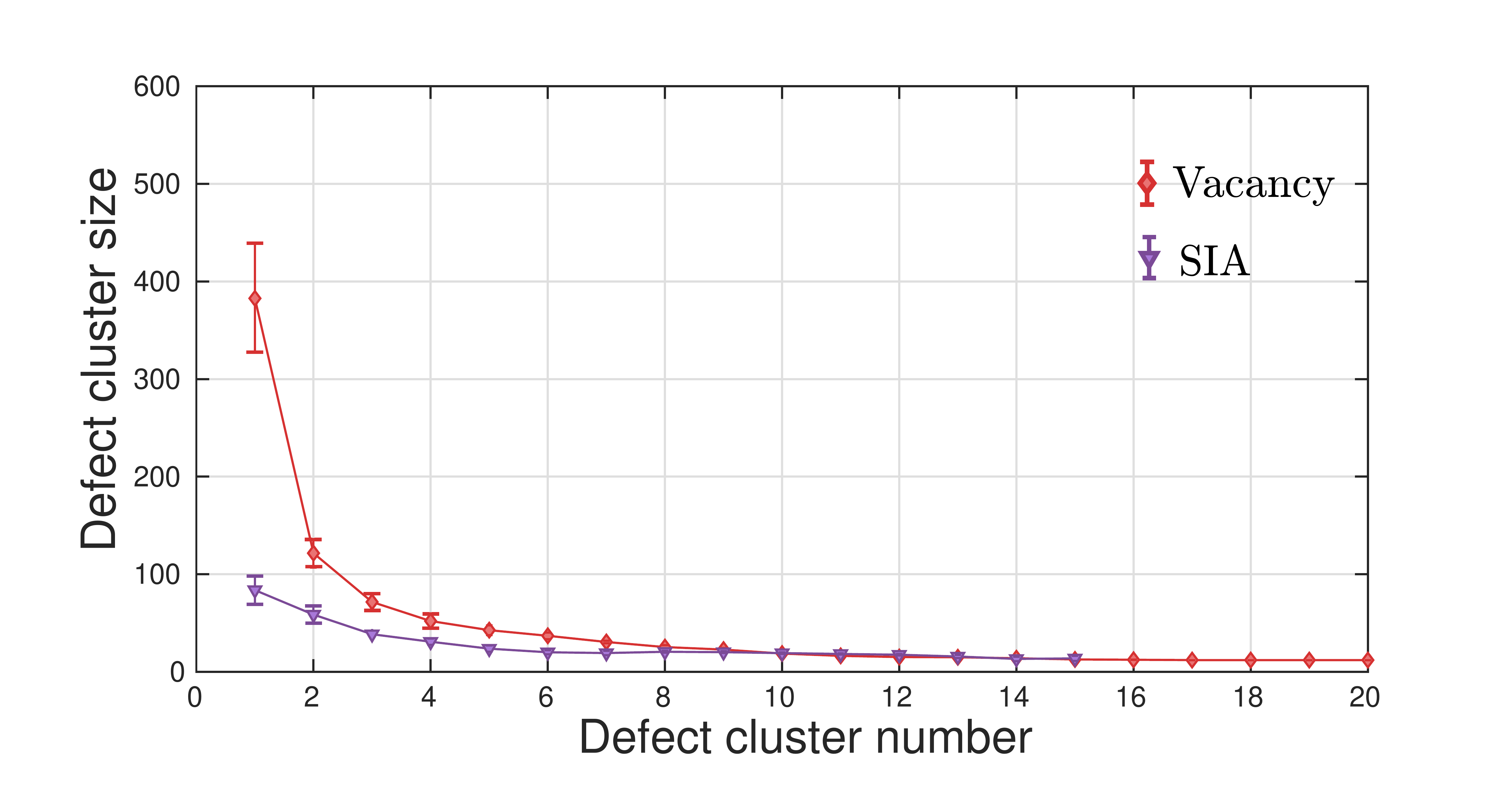}
         \caption{Nano-crystalline cell.}
         \label{Fig:cluster_count_nano}
     \end{subfigure}
        \caption{Defect clusters size for the biggest clusters in single crystal and nano-crystalline simulation cells for $50$ keV PKA energy showing that after the primary cascade event, the size of vacancy clusters was nearly unaffected while the SIA clusters were significantly reduced in size.}
        \label{Fig:cluster_count_nano_single}
\end{figure}

\subsection{Long-term diffusion using MXE method} \label{Section:MXE}
After simulating the primary radiation damage using $\ell$2T-MD, the long-term behavior of defects was analyzed with the MXE method for the NC damaged with $50$keV. The role of the GBs in interacting with the radiation-induced defects is presented next.  

\begin{figure} [H]
\centering
\includegraphics[width=1\textwidth]{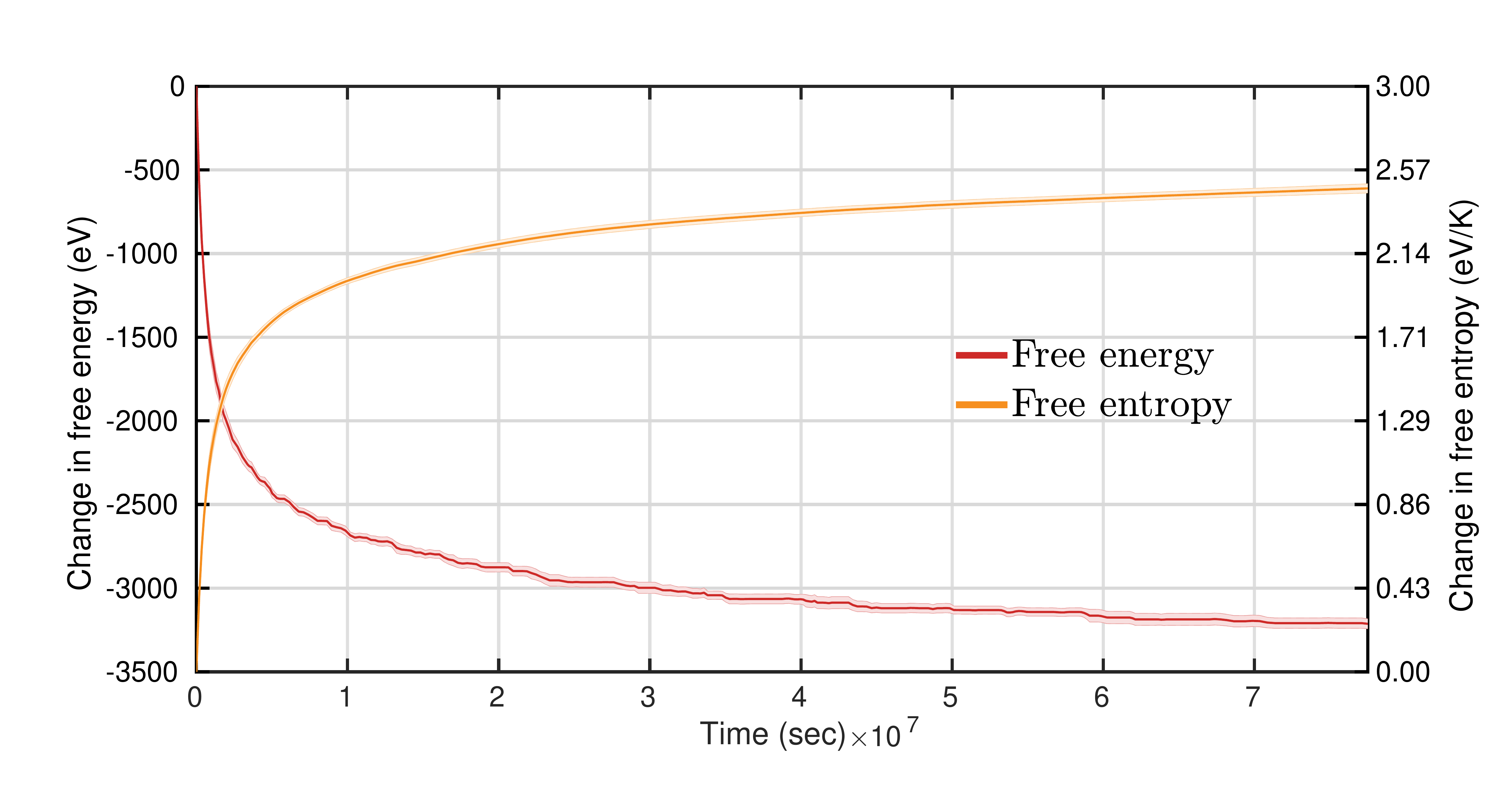}
\caption{Free entropy and free energy evolution with time, showing an increase of the change of free entropy relative to the initial value as the time progressed. Free energy of the system is minimized as diffusion of defects was allowed. }
\label{Fig:free_entropy}
\end{figure}

The time evolution of the free entropy and free energy calculated within the MXE framework using Eq. \ref{Eq:Grand-Canonical-Free-Energy} is shown in Fig.~\ref{Fig:free_entropy}. MXE minimizes the system's free energy (or, equivalently, increases the entropy). We observed that initially the rate of change of the free-energy was much higher, to quickly decline. This demeanor indicates that most of the diffusive events happened during that initial interval. 

\begin{figure} [H]
     \centering
     \begin{subfigure}[b]{0.48\textwidth}
         \centering
         \includegraphics[width=\textwidth]{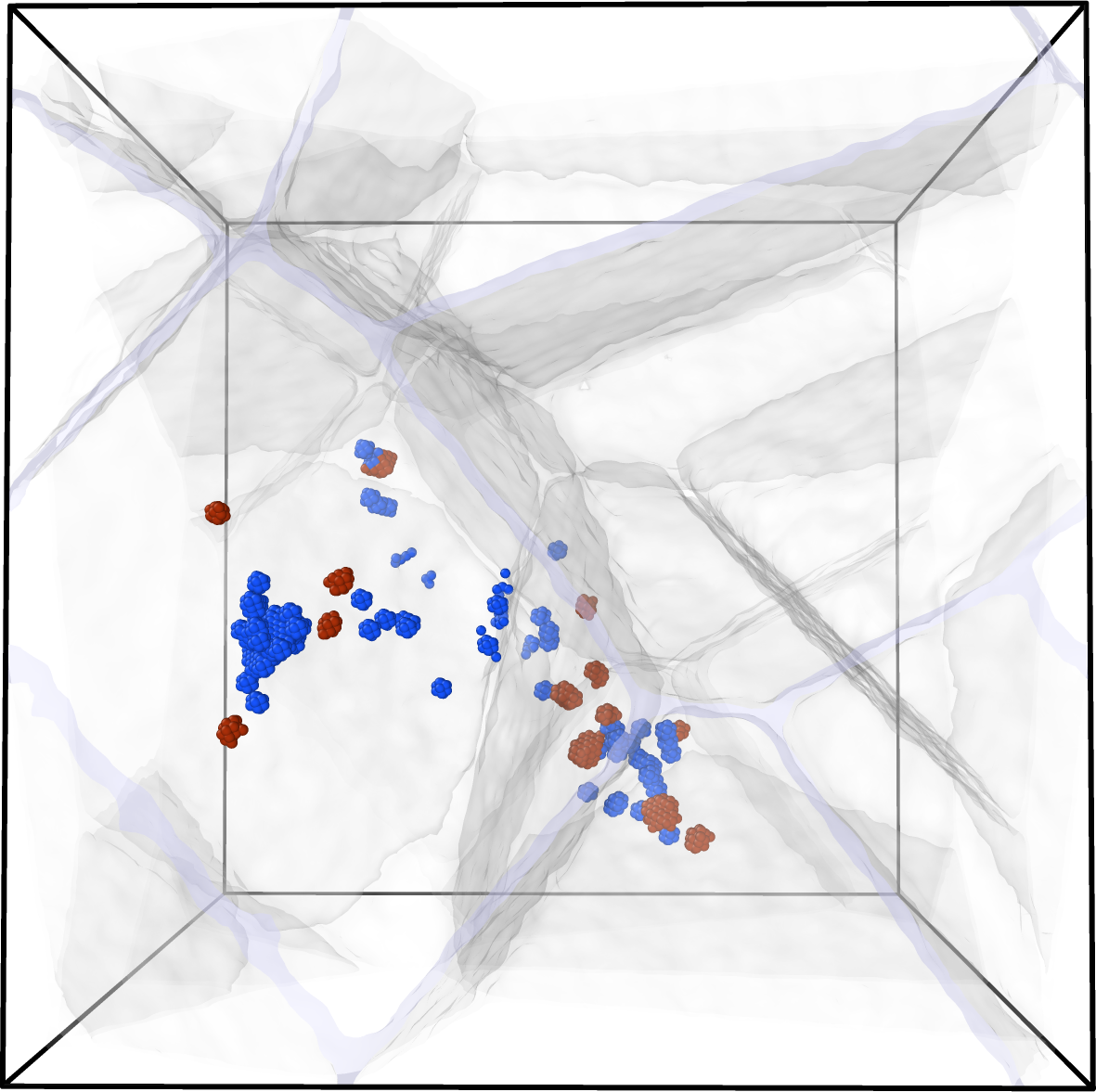}
         \caption{After cascade simulation with a 500 keV PKA.}
         \label{Fig:11a}
     \end{subfigure}
     \hfill
     \begin{subfigure}[b]{0.48\textwidth}
         \centering
         \includegraphics[width=\textwidth]{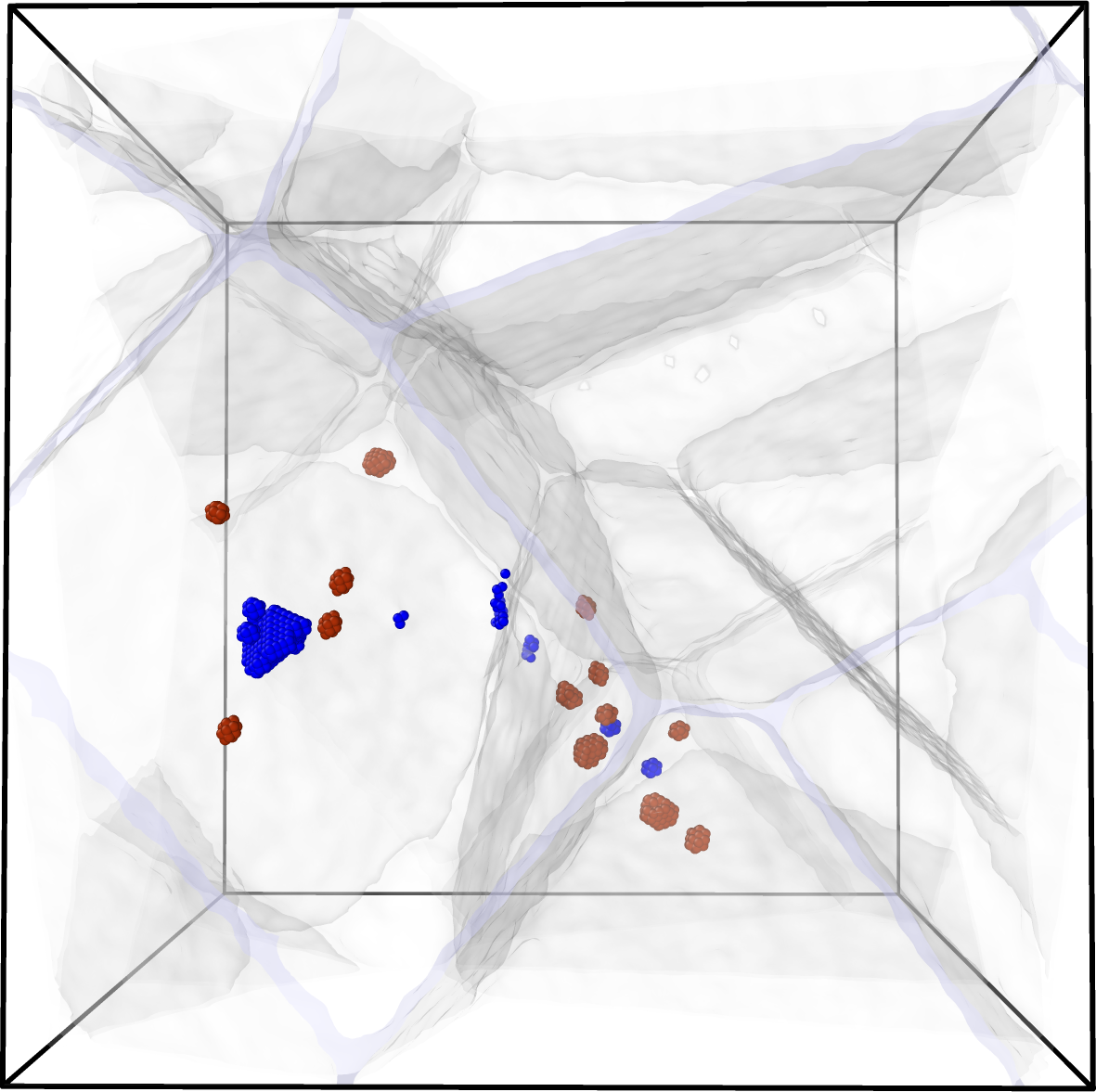}
         \caption{After defects diffusion.}
         \label{Fig:11b}
     \end{subfigure}
        \caption{
        Spatial distribution of the defect clusters in the NC simulation cell after (a) the cascade simulation using a $50$ keV PKA and (b) after defects diffusion with MXE. Blue clusters represents vacancy-based defects, red clusters represents SIAs-based defects. The grain boundary network of the nano-crystalline samples is shown as a transparent surface network.}  \label{Fig:vacancies_l2t_mxe}
\end{figure}

The defected clusters spatial configuration inside the material before and after the MXE simulation is shown Fig.~\ref{Fig:vacancies_l2t_mxe}. A qualitative comparison indicated a significant decrease in the number of vacancies in the bulk material during the long-term diffusion phase compared to the initial radiation damage stage (cf. Fig.~\ref{Fig:vacancies_l2t_mxe}(a) and Fig.~\ref{Fig:vacancies_l2t_mxe}(b)). It is evident from the graphical representation that the number of vacancies was reduced significantly while the SIA clusters showed a much slight reduction {\color{black}as shown in table  \ref{tab:Defect_table}}. 

\begin{figure} [H]
     \centering
     \begin{subfigure}[b]{0.8\textwidth}
         \centering
         \includegraphics[width=\textwidth]{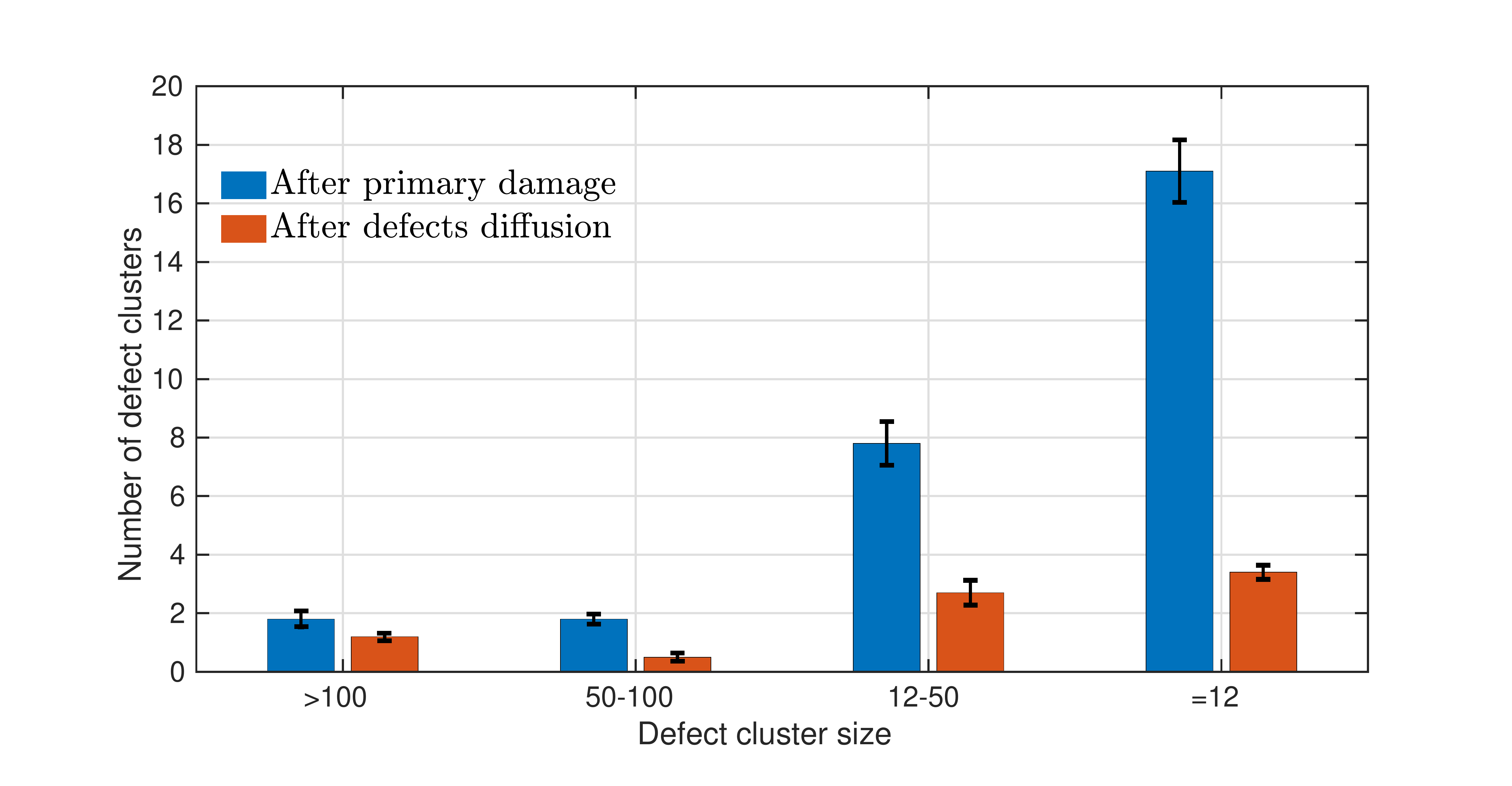}
         \caption{Vacancy clusters}
         \label{Fig:histo_cluster_l2t_mxe}
     \end{subfigure}
     \hfill
     \begin{subfigure}[b]{0.8\textwidth}
         \centering
         \includegraphics[width=\textwidth]{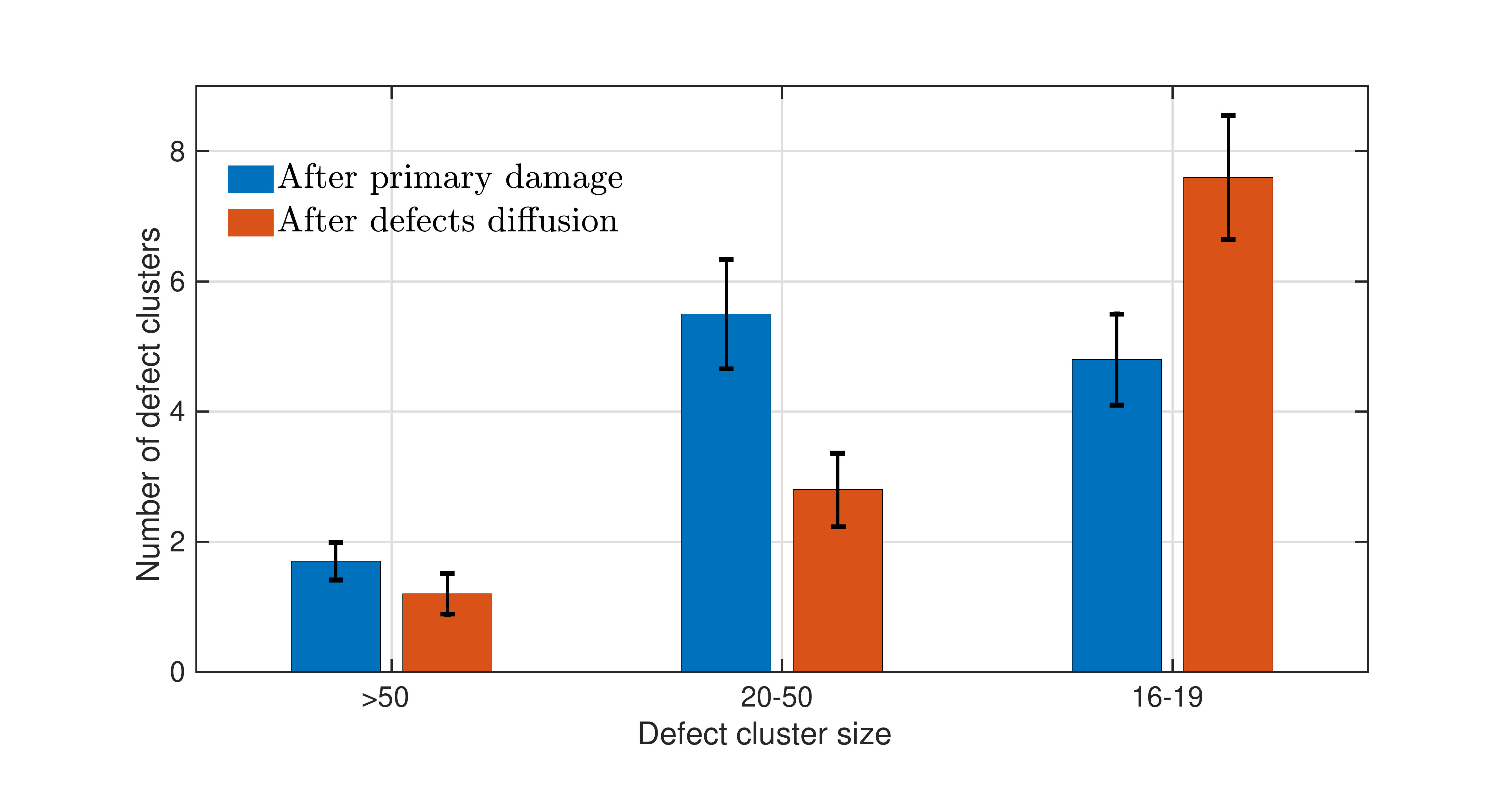}
         \caption{Self-Interstitial atoms clusters}
         \label{Fig:500_nano_l2t_mxe}
     \end{subfigure}
        \caption{Distribution of the number of defect clusters in a NC simulation cell after the cascade simulation using $\ell$2T-MD for $50$ keV PKA and after defects diffusion simulated with MXE for (a) vacancy clusters and (b) SIA clusters. Cluster size of $N_c=12$ represents the cluster of atoms around a single vacancy, while larger sizes represent clusters of more than one vacancy, while a cluster size of $16-19$ for SIA corresponds to one SIA.}
        \label{Fig:cluster_count_mxe_l2t}
\end{figure}

\begin{table*}[t]
\renewcommand{\arraystretch}{1.5} \setlength{\tabcolsep}{12pt}
\begin{center}
\begin{tabular}{cccccccc} 
\hline \hline
   &   & Largest defect cluster &  Number of defective atoms & Number of defect clusters   \\ \cline{3-5} 
\multirow{ 3}{*}{Vacancy} & single & 402 (106)  &  1162 (112)  & 33 (1.3) \\

& nano  & 382 (55) & 1069 (63) & 29 (1.1)    \\
& mxe &  284 (37) & 474 (41) & 8 (0.5)    \\[0.25cm]
\hline

\multirow{ 3}{*}{SIA} & single & 272 (46)  &  1062 (60) & 23 (1.8)    \\
& nano  & 84 (15) & 406 (47) & 13 (1.4)   \\
& mxe &  73 (14) & 337 (39) & 12 (1.1) \\


 \hline \hline
\end{tabular}
\end{center}
\caption{The mean values and standard error (in brackets) for the largest defect cluster, which is the number of atoms forming the largest defect cluster. The number of defective atoms, which is the total number of non-FCC atoms. The number of defect clusters, which is the count of the defect clusters. }
\label{tab:Defect_table}
\end{table*}

Let us now focus on quantifying the variation of the defects before and after the MXE simulation. Figure \ref{Fig:cluster_count_mxe_l2t} shows the evolution of the clusters before and after the MXE simulation {\color{black}for vacancy and SIA defect clusters}. Focusing on Fig.~\ref{Fig:histo_cluster_l2t_mxe}, where the number of defect clusters \emph{vs.} cluster size ({\color{black}$N_c$}) is shown before, and after the MXE simulation, we observed that the number of {\color{black}vacancy} clusters {\color{black}had been reduced for all the sizes} after the MXE simulation. For instance, {\color{black}individual} vacancies ($N_c=12$) were reduced from {\color{black}$17$ to less than $4$} after the MXE simulation, representing a reduction of {\color{black}$\sim76\%$} of all vacancies. Similar behaviour was found for other clusters with sizes between $13-50$ atoms, i.e., from {\color{black} about $8$ to $3$ for clusters with $13-50$ atoms (about $62\%$ reduction).} {\color{black}For the SIA clusters, a similar behaviour was observed except for the individual SIA clusters with $16-19$, which their number increased in expense of the larger clusters.} 

\begin{figure} [H]
     \centering
    \begin{subfigure}[b]{0.46\textwidth}
         \centering
         \includegraphics[width=\textwidth]{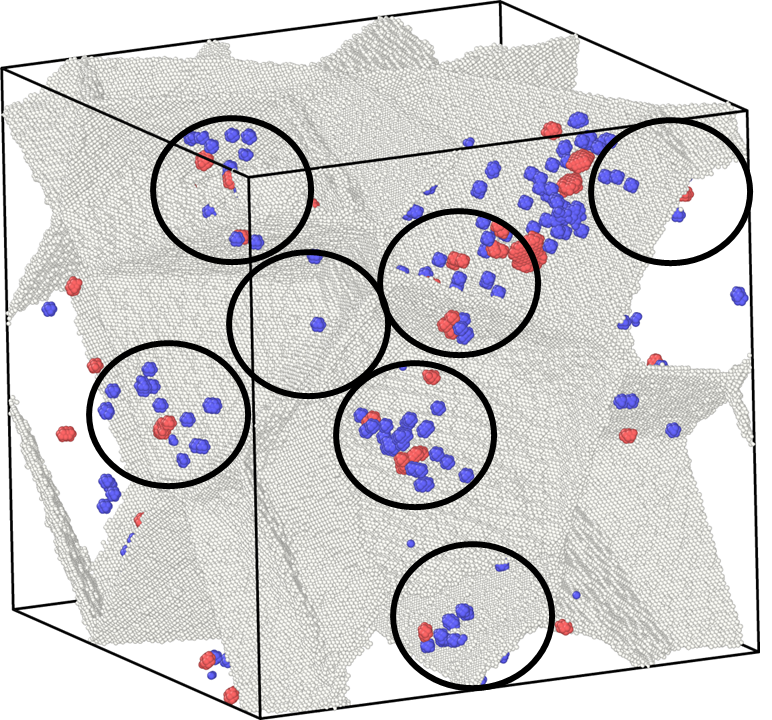}
         \caption{GB network and defect clusters before MXE.}
         \label{Fig:14a}
     \end{subfigure}
     \begin{subfigure}[b]{0.46\textwidth}
         \centering
         \includegraphics[width=\textwidth]{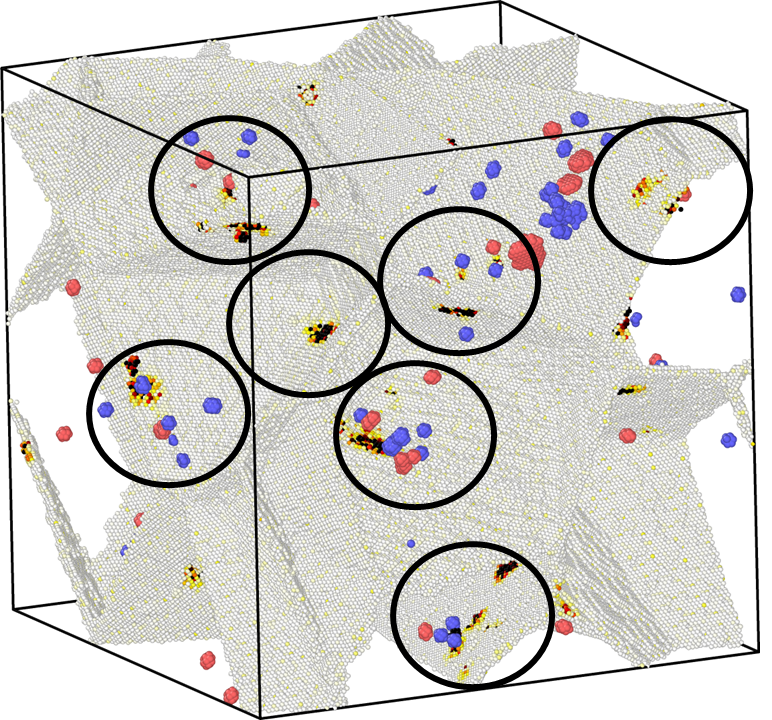}
         \caption{GB network and defect clusters after MXE.}
         \label{Fig:14b}
     \end{subfigure}
     
        \caption{Comparison of the defect clusters before and after the MXE simulation. (a) GB network with the defect clusters after the radiation damage simulation. 
        Blue atoms represent vacancies and red  atoms represent interstitials. Several areas of interest were highlighted with circles.
         (b) GB network and defect cluster after the MXE simulation. Sites in the GB responsible for absorbing the vacancies and interstitials are also shown in a gradient of color from yellow to black represents the sites change in atomic molar fraction.} 
        \label{Fig:void_evolution}
\end{figure}

Let us now present the spatial and temporal evolution of the defects in the NC sample. The evolution of defects from the bulk of the nano-grains to the GBs can be seen in Fig.~\ref{Fig:void_evolution}, where the GB network and defect clusters were plotted before (see Fig.~\ref{Fig:14a}) and after the MXE simulation (see Fig.~\ref{Fig:14b}). An animated evolution of these defects is shown in the {\bf Supplementary Video 3} for reference. As time went by, the defect clusters tended to diffuse to the GB network. By tracking the change in the site's atomic molar fraction ($x_i$) in the GB network, we can identify those sites that are more likely to annihilate the vacancies from the bulk. Remarkably, only certain regions are responsible for annihilating the vacancies in the GB network (circled in Fig.~\ref{Fig:void_evolution}). These regions are determined by several factors, including the distance between the GB and defect, chemical potential, and the geometrical arrangement of the sites.  

\begin{figure} [H]
\centering
\includegraphics[width=1\textwidth]{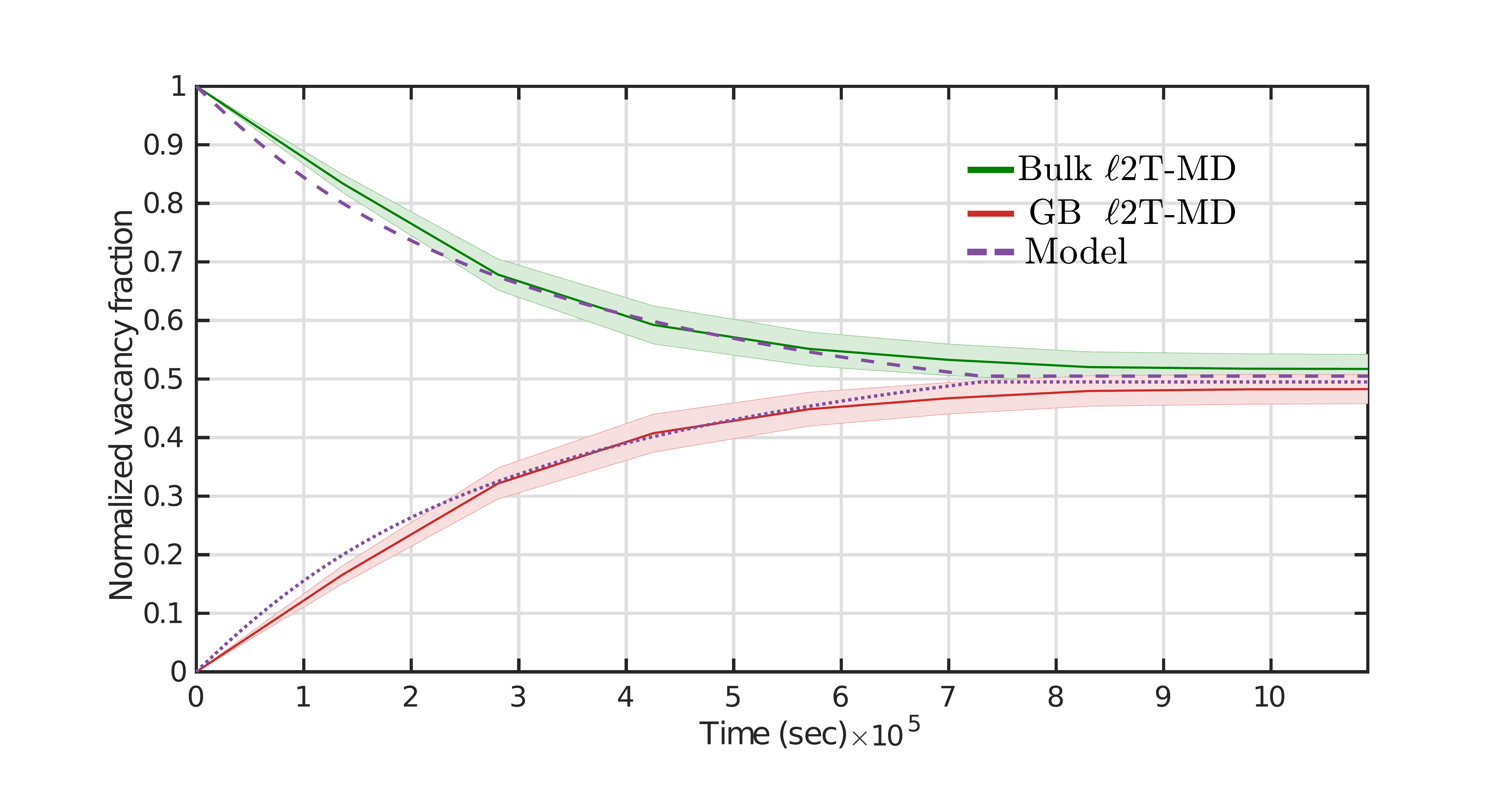}
\caption{The evolution of the fraction of vacancies in bulk and GB with time showing the equilibrium concentration for bulk and GB. The dashed and dotted lines represent the results obtained with a phenomenological model discussed in Section \ref{Section:Discussions}. }
\label{Fig:15}
\end{figure}

To better understand the mass exchange between the bulk of the nano-grains and the GB network, we plotted in Fig. \ref{Fig:15} the time evolution of the vacancies during the MXE simulation for the bulk and GB of the cell. Initially, all vacancies were located inside the nano-grains, and the vacancies in the GB were zero. However, as diffusion was allowed in the sample, vacancies migrated from the bulk to the GB network. Only about {\color{black}50\%} of the total vacancies in the nano-grains survived after the MXE simulation since vacancies were annihilated in the GB network. Of tremendous interest, the time evolution of the vacancies indicated a minimum of vacancies in the cell, and diffusion of mass stopped at around {\color{black}$\sim8 \times 10^5$} s. This effect is attributed to the collective behaviour of vacancy (or cluster of vacancies). Since the GB are amorphous, vacancies can migrate to the GBs and remove residual stresses, reducing the chemical potential gradient and the system's free energy. However, this effect saturates when the chemical potential is in equilibrium. Hence, diffusion of vacancies reached a steady-state, as shown in Fig.~\ref{Fig:15} at the end of the simulation. A visual inspection of Fig.~\ref{Fig:void_evolution} indicates that the remaining vacancies were more or less clustered in the sample, indicating much larger defects than individual vacancy sites, which, ultimately, slows down the diffusion of these surviving clusters.

\begin{figure} 
     \centering
      \begin{subfigure}[b]{0.4\textwidth}
         \centering
         \includegraphics[width=\textwidth]{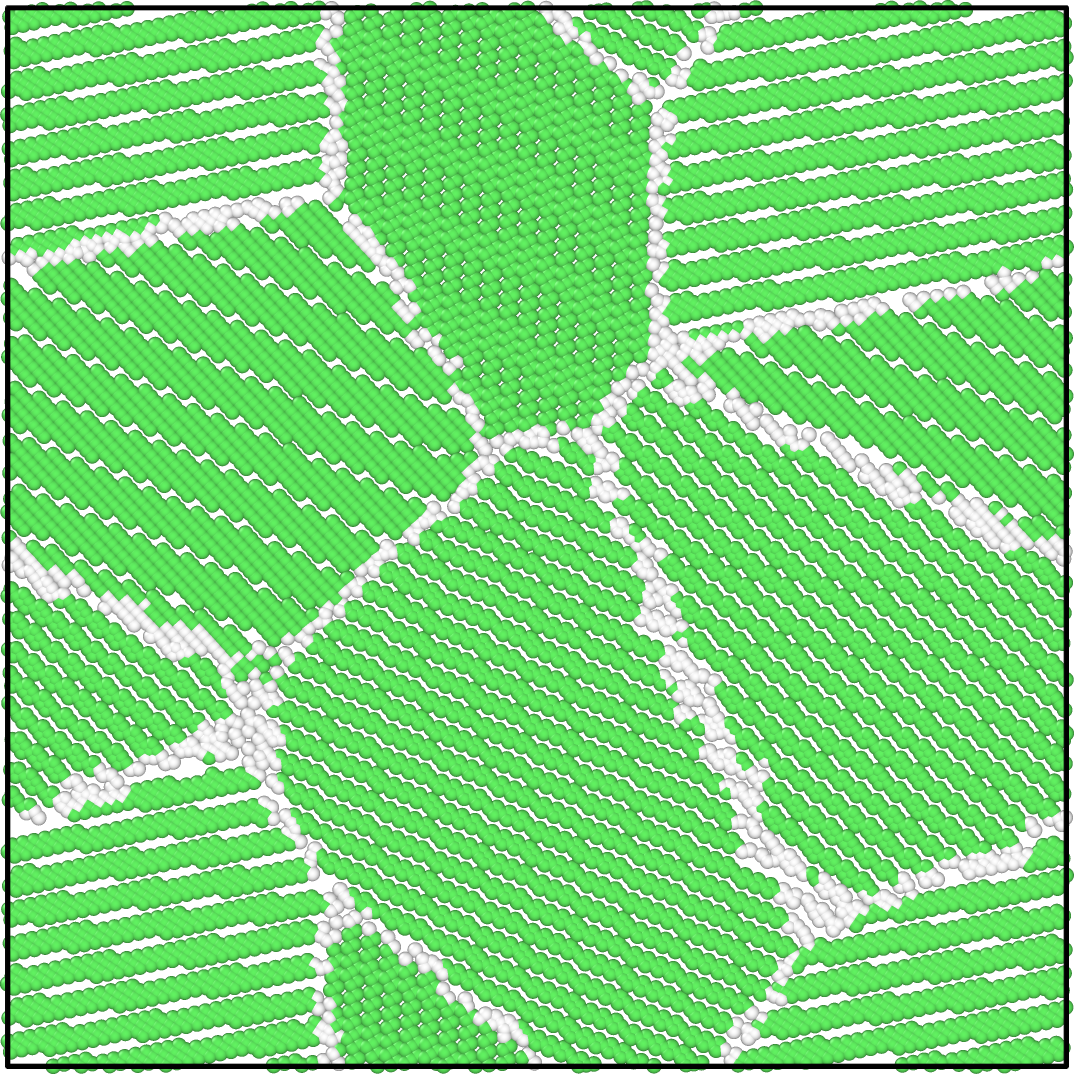}
         \caption{GB atoms in white and bulk atoms in green}
         \label{Fig:16a}
     \end{subfigure}
     \begin{subfigure}[b]{0.4\textwidth}
         \centering
         \includegraphics[width=\textwidth]{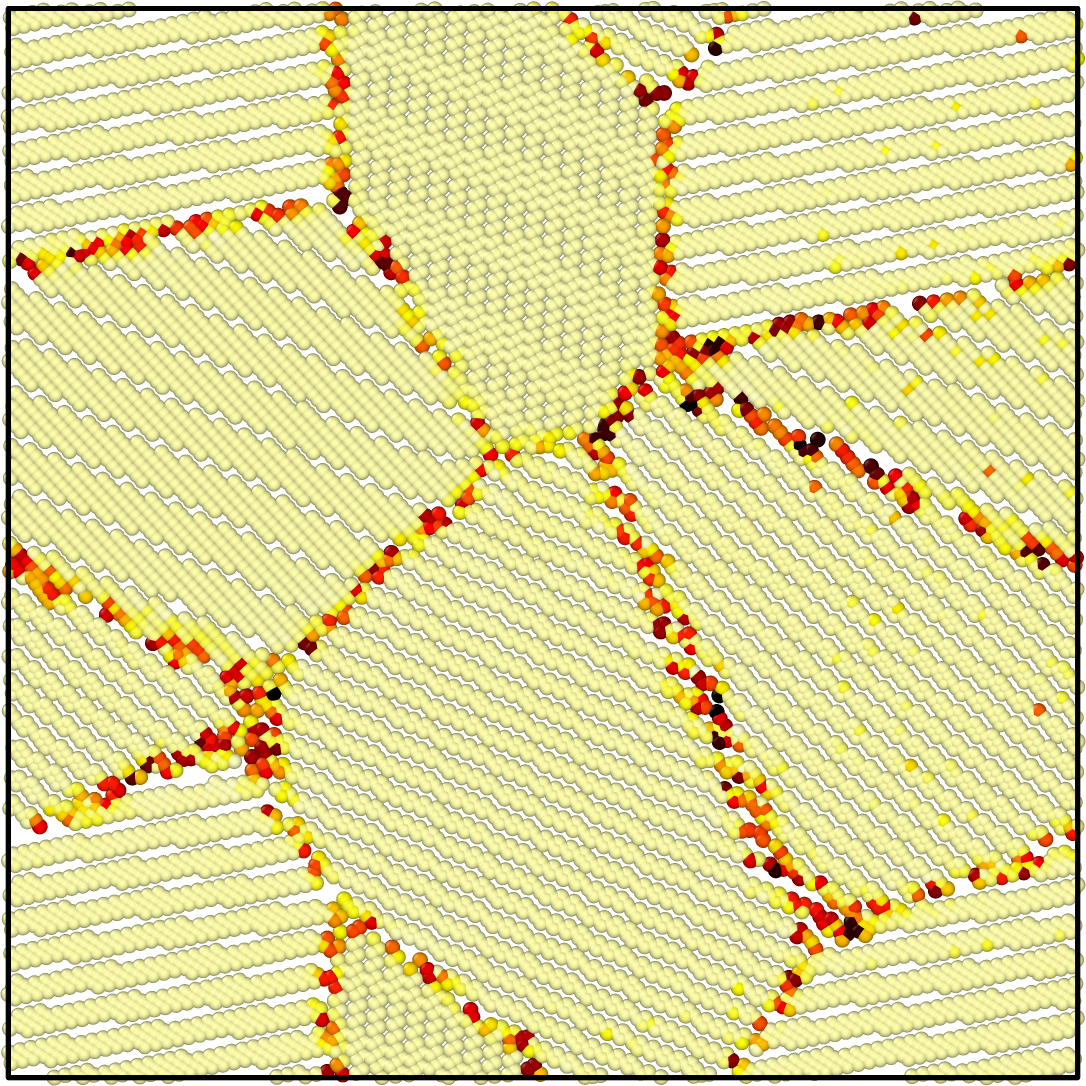}
         \caption{Vacancy formation energy distribution}
         \label{Fig:16b}
     \end{subfigure}

    \begin{subfigure}[b]{0.25\textwidth}
         \centering
         \includegraphics[width=\textwidth]{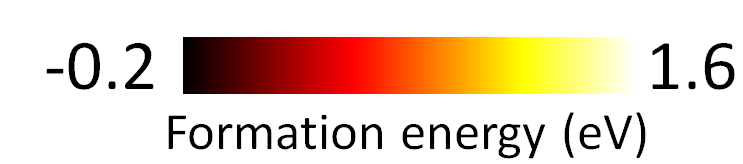}
         \label{Fig:formation_color}
     \end{subfigure}
     
     \begin{subfigure}[b]{0.45\textwidth}
         \centering
         \includegraphics[width=\textwidth]{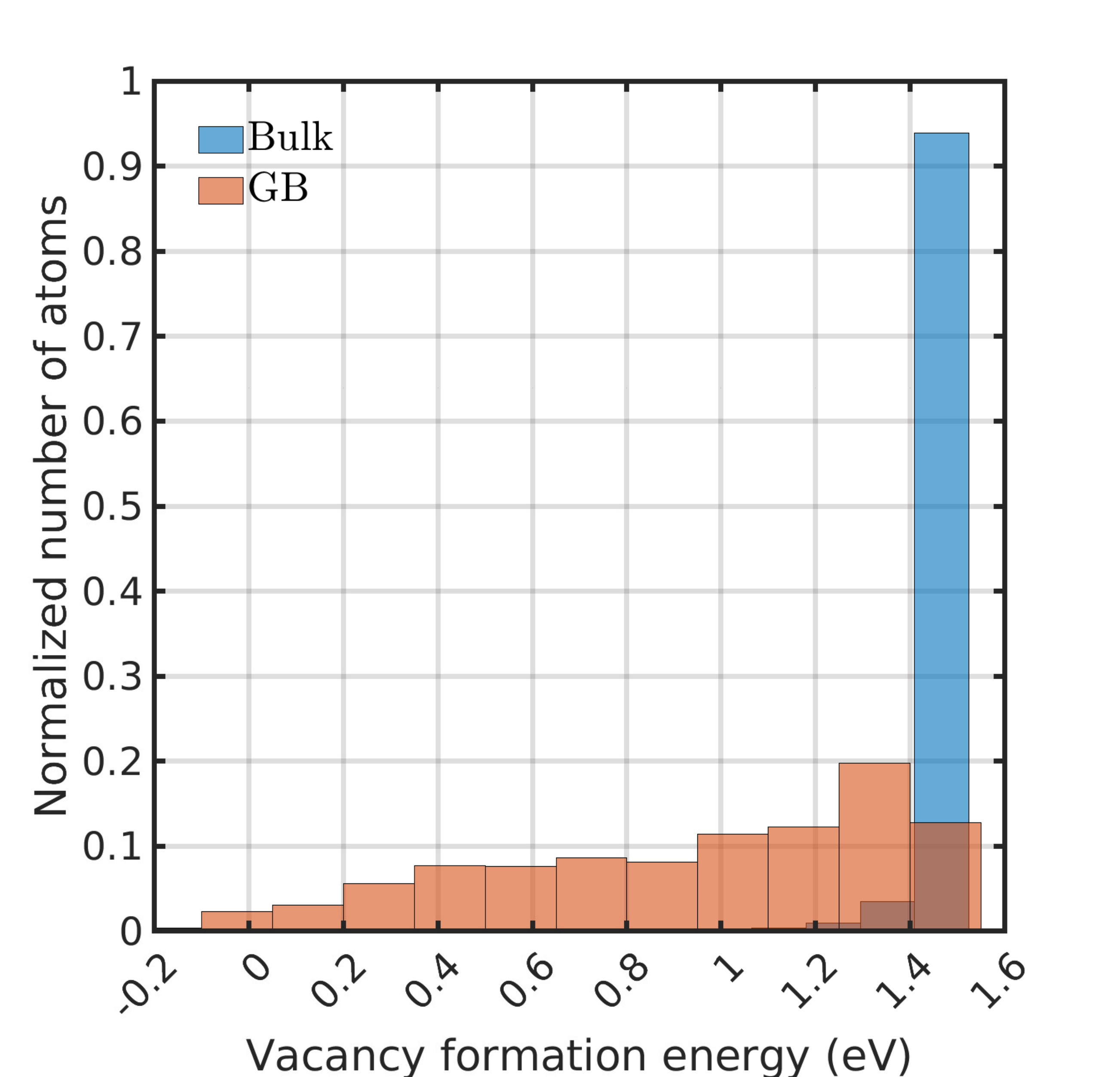}
         \caption{Vacancy formation energy for GB and bulk}
         \label{Fig:16c}
     \end{subfigure}
     \begin{subfigure}[b]{0.45\textwidth}
         \centering
         \includegraphics[width=\textwidth]{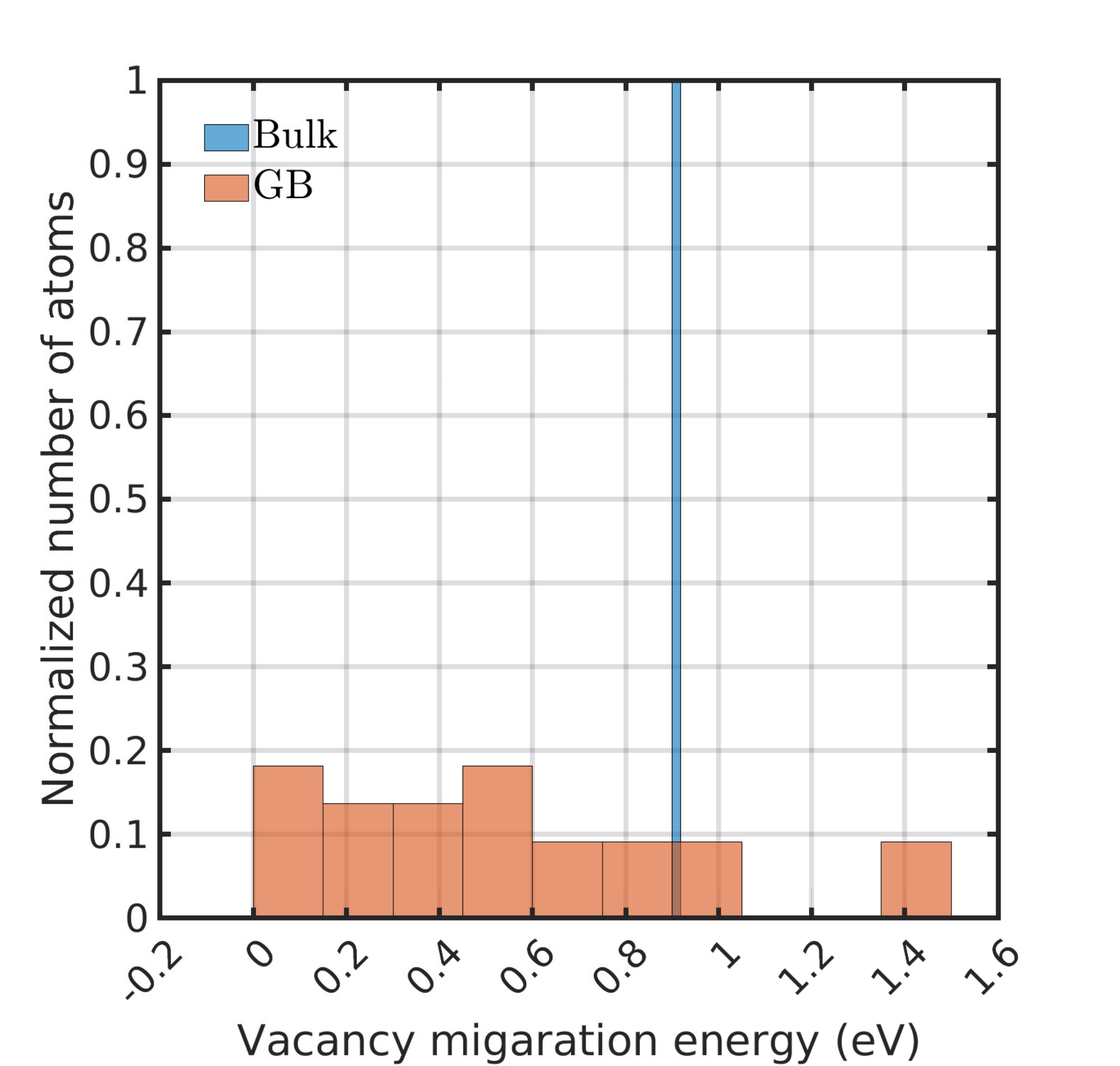}
         \caption{Vacancy migration energy for GB and bulk}
         \label{Fig:16d}
     \end{subfigure}
        \caption{Vacancy formation ($Q_f$) and migration ($Q_m$) energies distributions indicated most GB sites have lower energies than sites placed in the bulk of the NC. This lower energies suggest a tendency for vacancies to segregate and migrate inside GBs. (a) A slice of the simulation cell with a thickness of $a_0$ showing GB atoms in white and bulk atoms in green (b) Vacancy formation energy for every atom in the same slab as in (a). (c) and (d) show the Vacancy formation and migration energy distribution densities in the slab.}
        \label{Fig:16}
\end{figure}

To elucidate the effect of the GB and their interaction with vacancies, we investigated the formation and migration energy of a slab with thickness $b$ of the NC sample, shown in Fig.~\ref{Fig:16}. Figure \ref{Fig:16a} shows the arrangement of the atoms in the sample with atoms in the bulk (green) and the GBs (white), while Fig.~\ref{Fig:16b} shows the same slab but with atoms colored according to their formation energy. Besides some local fluctuations, the formation energy in the bulk is pretty homogeneous. The vacancy formation energy in the bulk was around $\sim1.5$ eV (see Fig.~\ref{Fig:16c}), with minor fluctuations in the sample due to the internal stresses that appeared due to the GBs. These values are in close agreement to previously published results \cite{garcia2002self}. At the GBs, however, the formation energy had a large spread, from $0-1.55$ eV, with the spatial distribution shown in Fig.~\ref{Fig:16b} and the probability distribution function shown in Fig.~\ref{Fig:16c}. The analysis revealed that most of the sites in the GBs are energetically more favorable to create vacancies than the bulk. However, for some isolated sites, the formation energy was slightly higher than the bulk, suggesting that they are not favorable to annihilated vacancies. 

To complete this picture, we also investigated the migration energy for the sites in the slab, as shown in Fig.~\ref{Fig:16d}. In the bulk, the migration energy was pretty stable and close to $\sim0.9$ eV. On the other hand, for sites located in the GBs, the migration energy had a much more significant fluctuation. Most sites had migration energy oscillating between $0-1.0$ eV with a flat distribution, as shown in Fig.~\ref{Fig:16d}. Only a small percentage of atoms had much higher migration energy, i.e., $\sim1.4$ eV, representing about 10\% of all sites. Thus, the analysis indicated two effects that combine to explain the superior healing properties of NC materials. First, vacancies are prone to be annihilated in the GBs due to reduced formation energies in these sites, and this process reduces the energy in the NC cell. Secondly, a much smaller barrier for the migration of those vacancies in the GB network allows vacancies to diffuse to the most energetically favorable sites quickly. However, not all sites were more favorable for the annihilation of vacancies, and some sites also showed much higher migration energy. These effects combined with the slower diffusion of the cluster of vacancies generate a saturation effect for the annihilation of vacancies in the GBs, explaining the long-term behavior of vacancies presented in Fig. \ref{Fig:15}.

\begin{figure} [H]
     \centering
     \begin{subfigure}[b]{0.32\textwidth}
         \centering
         \includegraphics[width=\textwidth]{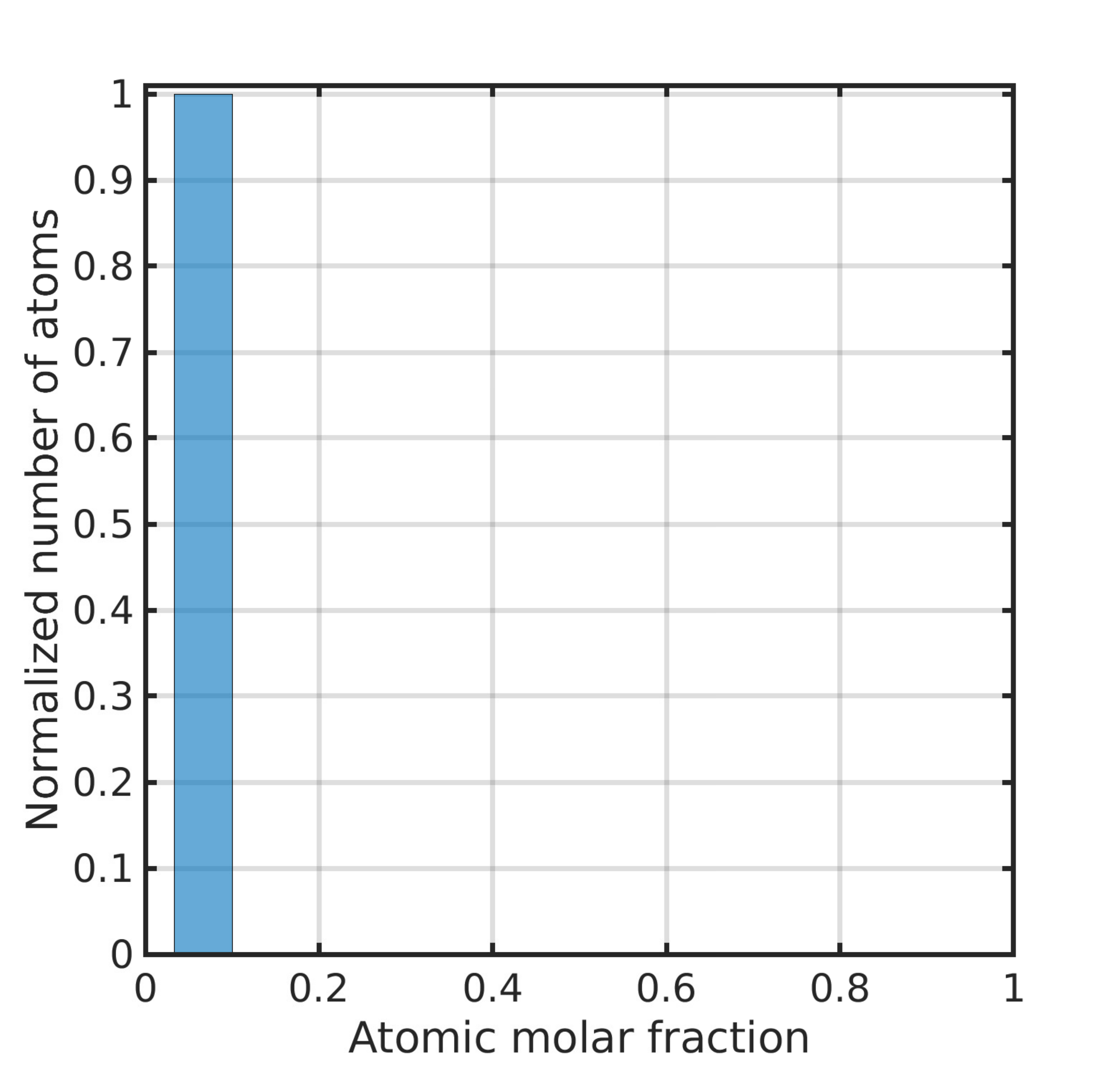}
         \caption{$0$ sec}
         \label{Fig:xi0}
     \end{subfigure}
     \begin{subfigure}[b]{0.32\textwidth}
         \centering
         \includegraphics[width=\textwidth]{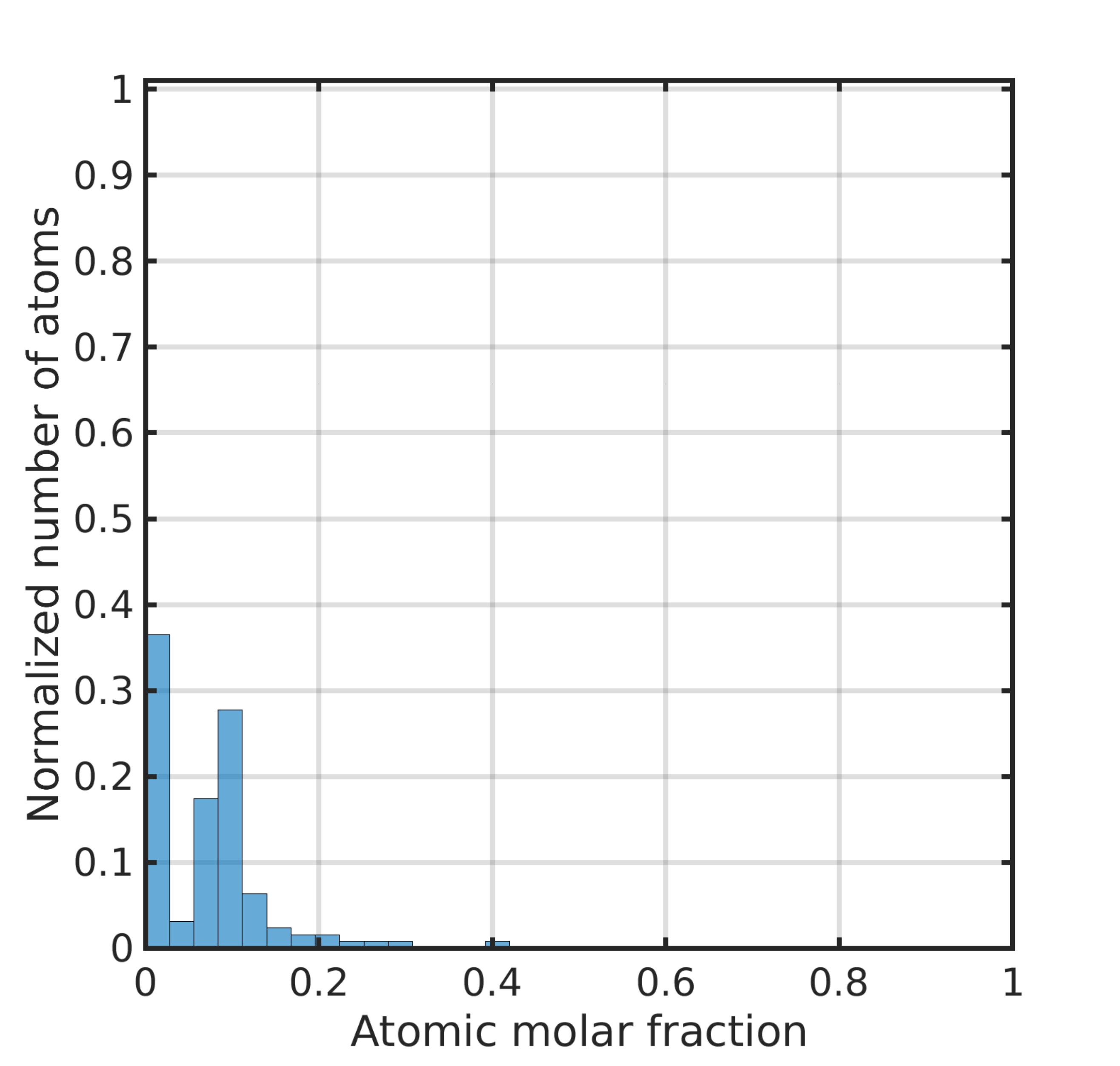}
         \caption{$1.84\times10^{4}$ sec}
         \label{Fig:xi1}
     \end{subfigure}
          \begin{subfigure}[b]{0.32\textwidth}
         \centering
         \includegraphics[width=\textwidth]{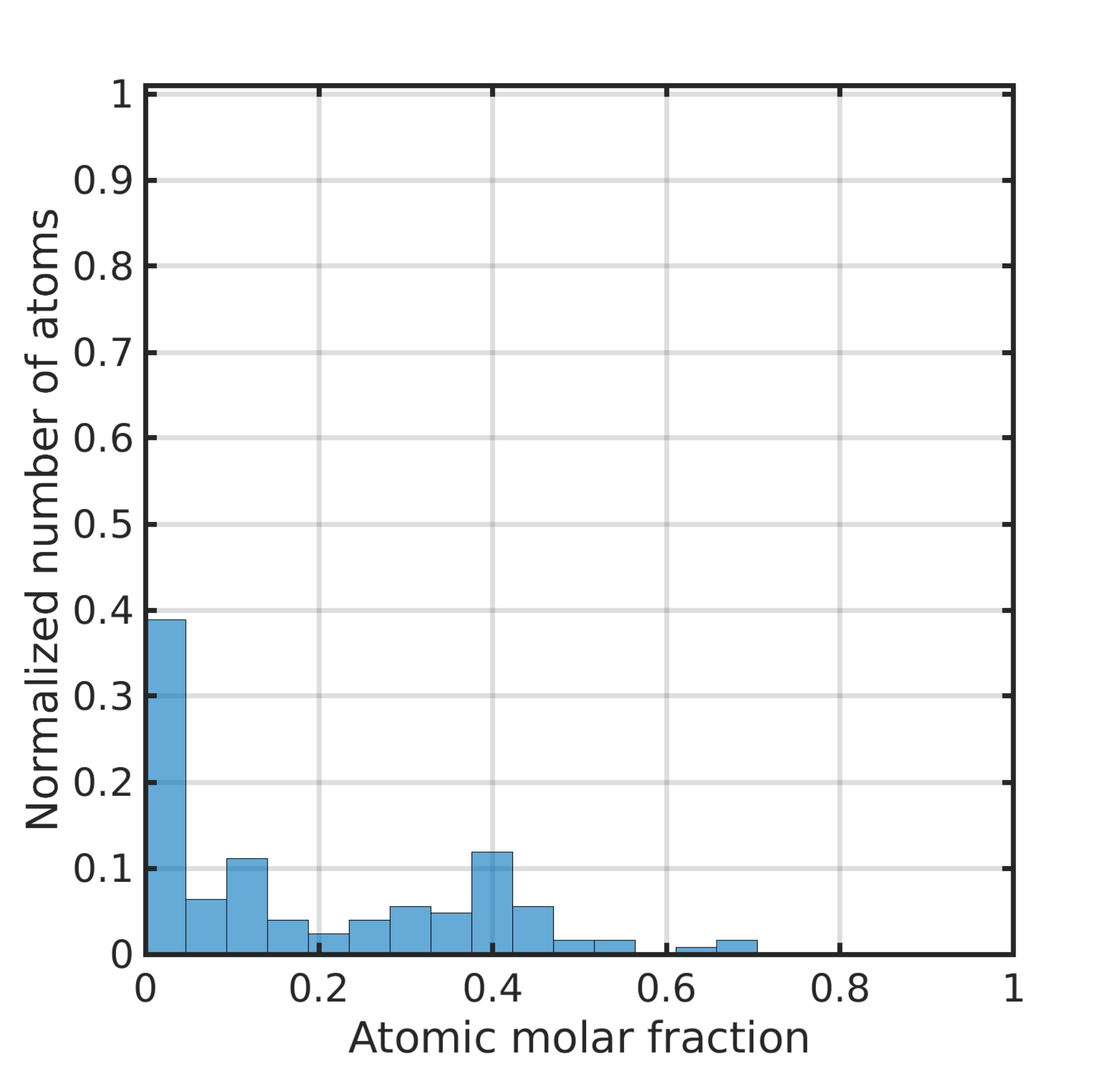}
         \caption{$3.43\times10^{5}$ sec}
         \label{Fig:xi10}
     \end{subfigure}
     \begin{subfigure}[b]{0.32\textwidth}
         \centering
         \includegraphics[width=\textwidth]{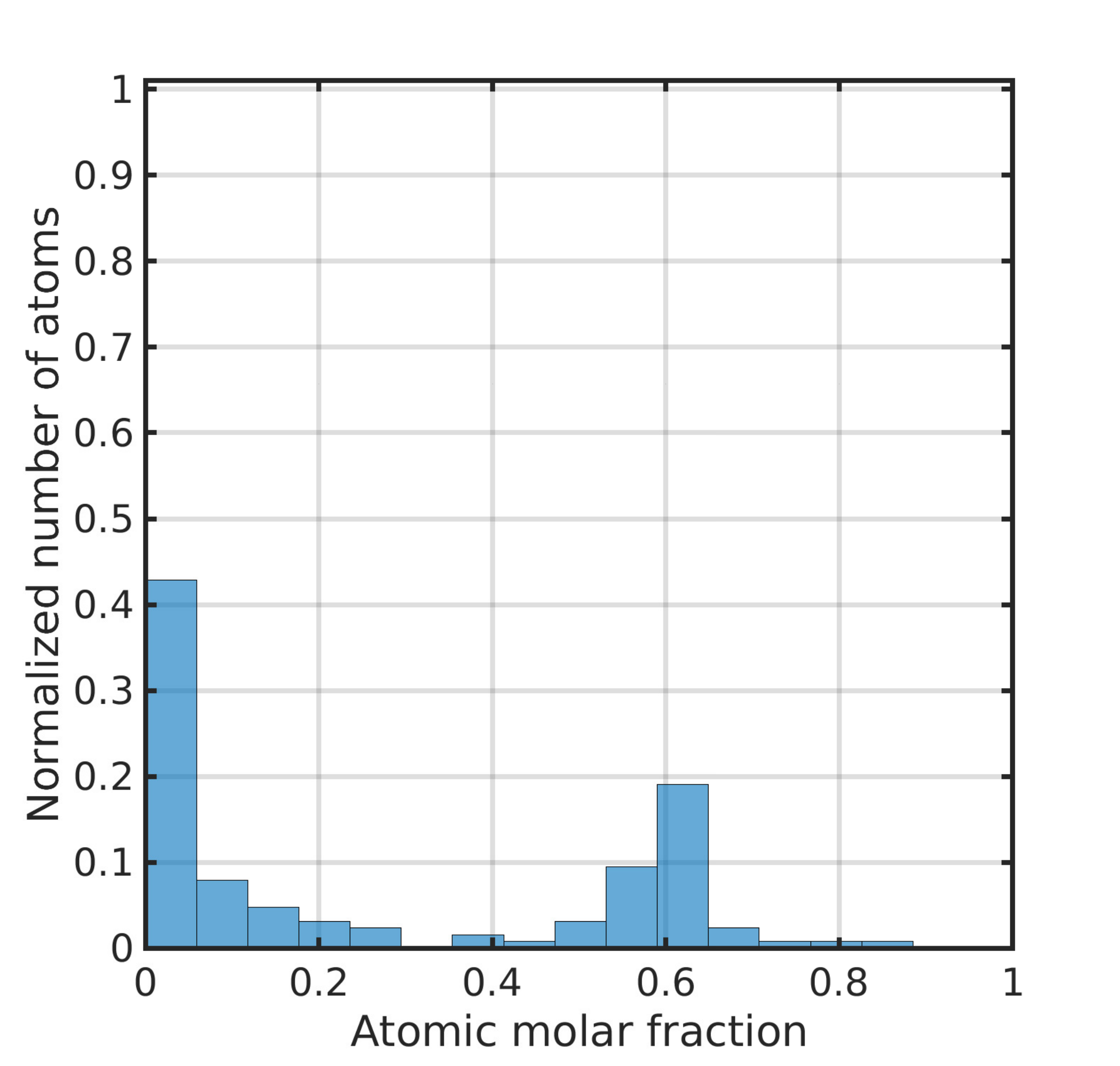}
         \caption{$4.98\times10^{6}$ sec}
         \label{Fig:xi15}
     \end{subfigure}
          \begin{subfigure}[b]{0.32\textwidth}
         \centering
         \includegraphics[width=\textwidth]{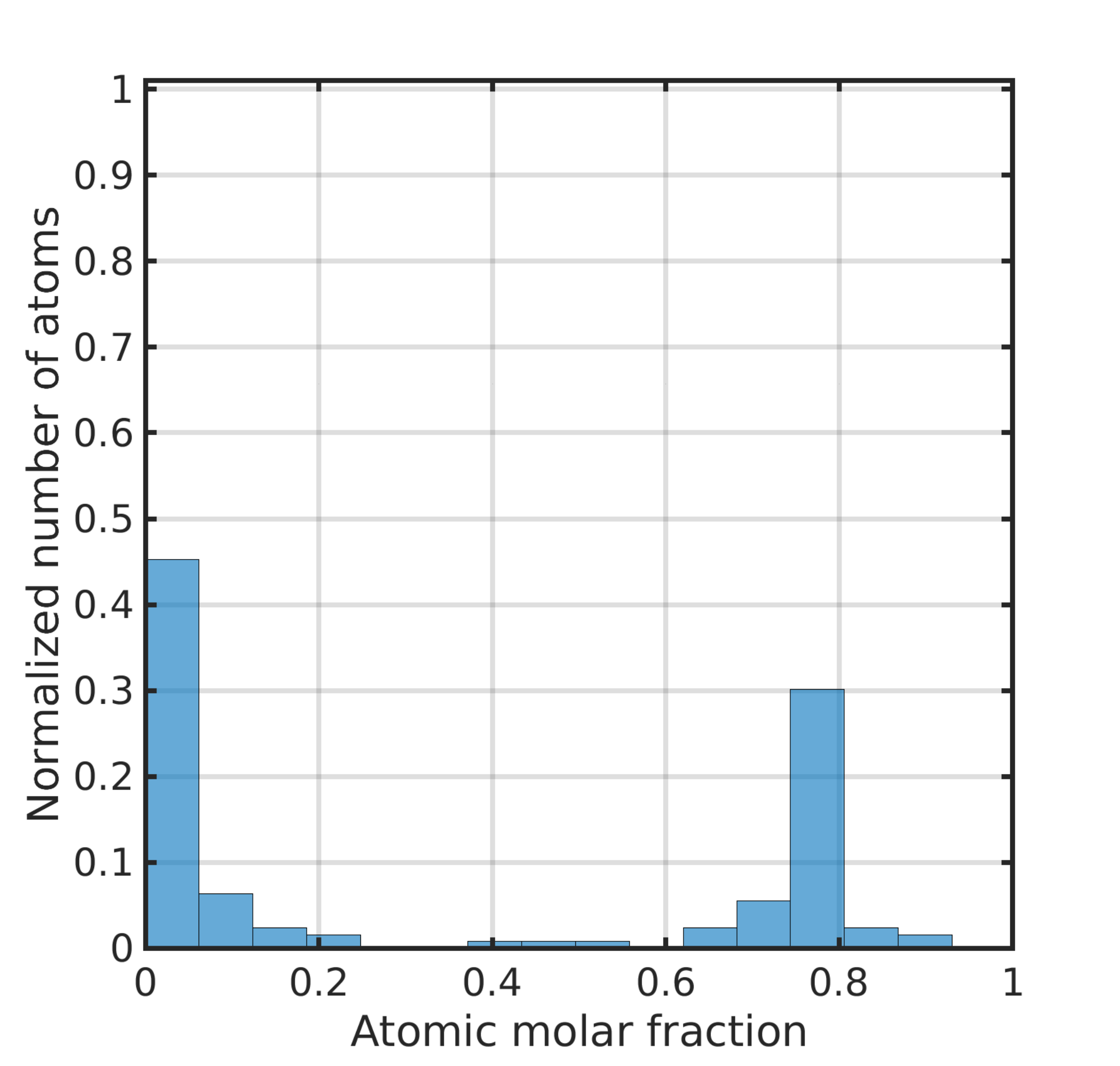}
         \caption{$6.52\times10^{6}$ sec}
         \label{Fig:xi20}
     \end{subfigure}
     \begin{subfigure}[b]{0.32\textwidth}
         \centering
         \includegraphics[width=\textwidth]{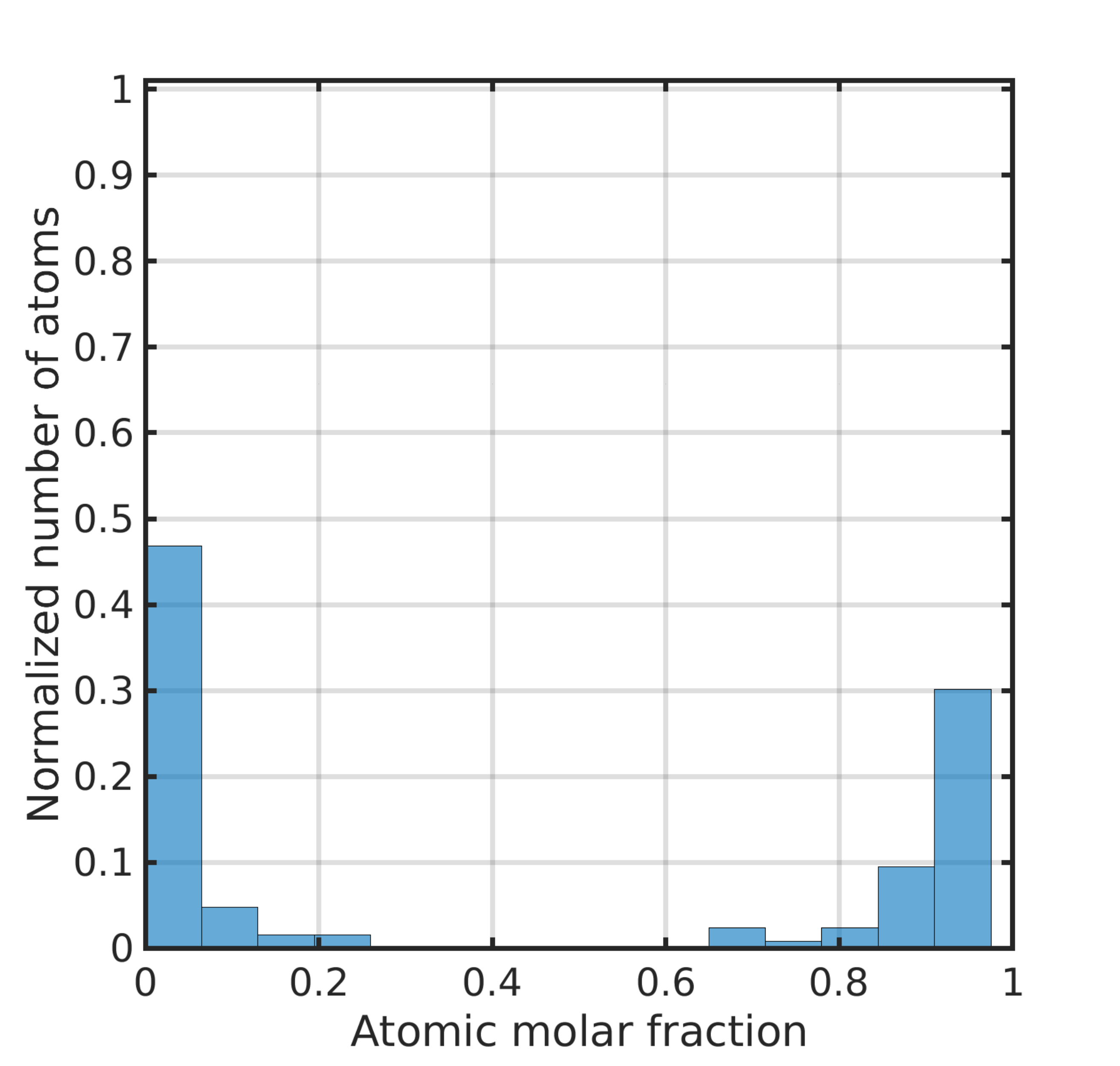}
         \caption{$2.34\times10^{7}$ sec}
         \label{Fig:xi102}
     \end{subfigure}
        \caption{Vacancy distribution density time evolution for the NC simulation cell during defects diffusion in MXE. (a) Initial distribution is represented by a delta Dirac function. (d)-(e) As time progressed, the distribution split in two, sites with stable vacancies, and sites fully occupied. (f) Final distribution with sites either fully empty ($\sim48\%$) or fully occupied.  }
        \label{Fig:Xi_evolution_2}
\end{figure}

After studying the evolution of individual defects in detail with the MXE simulation, we describe the macroscopic evolution of the vacancies. We tracked down the spatial and temporal evolution of all vacancies after the PKA primary damage. This evolution is important to understand the collective behavior of vacancies in NC samples and help understand their behavior and develop coarse-grained models for continuum scale methods. Figure \ref{Fig:Xi_evolution_2} shows the time evolution of the normalized number of vacancy atoms \emph{vs.} the atomic molar fraction. Initially, all vacancy sites were placed in the bulk of the NC at different locations from the GBs network, and their distribution was similar to a delta Dirac function, as shown in Fig.~\ref{Fig:Xi_evolution_2}(a). As diffusion was allowed during the MXE simulation, the atomic molar fraction for those atoms changed, as shown in the subsequent subfigures. Remarkably, after only 10\% of the total simulated time, a subset of the sites representing about $\sim40\%$ of the initial vacancies remained stable with very low atomic molar fractions ($x_i = 0-0.1$), as shown in Figs. \ref{Fig:Xi_evolution_2}(b) and (c). This group of sites represents the surviving vacancies in the simulation with very stable configurations. As a result, they showed very little change over time after that. The remaining portion of initial vacancy sites increased their concentration, as shown by the plots in Figs. \ref{Fig:Xi_evolution_2}(b)-(d). The atomic molar fraction's increment for the sites containing single vacancies indicated that they were not energetically favorable, and their position changed with time. Thus, the vacant sites ($x_i=0$) originally located in the bulk of the NC transformed into occupied lattice sites ($x_i=1$), and the vacancies were then homogeneously distributed across the GB network. This transition was not instantaneous, and a long time pass to obtain sites fully occupied. This behaviour is illustrated in Figs. \ref{Fig:Xi_evolution_2}(e)-(f) where most of the sites had either low ($x_i\le0.1$) or high ($x_i\ge0.95$) atomic molar fractions. We also observed that the velocity at which these sites increased their atomic molar fraction was not constant, and there was a distribution that depended on multiple factors, including the local chemical potential, the internal stresses, and the geometrical features of the GB network. 

\begin{figure} [H]
     \centering
     \begin{subfigure}[b]{0.32\textwidth}
         \centering
         \includegraphics[width=\textwidth]{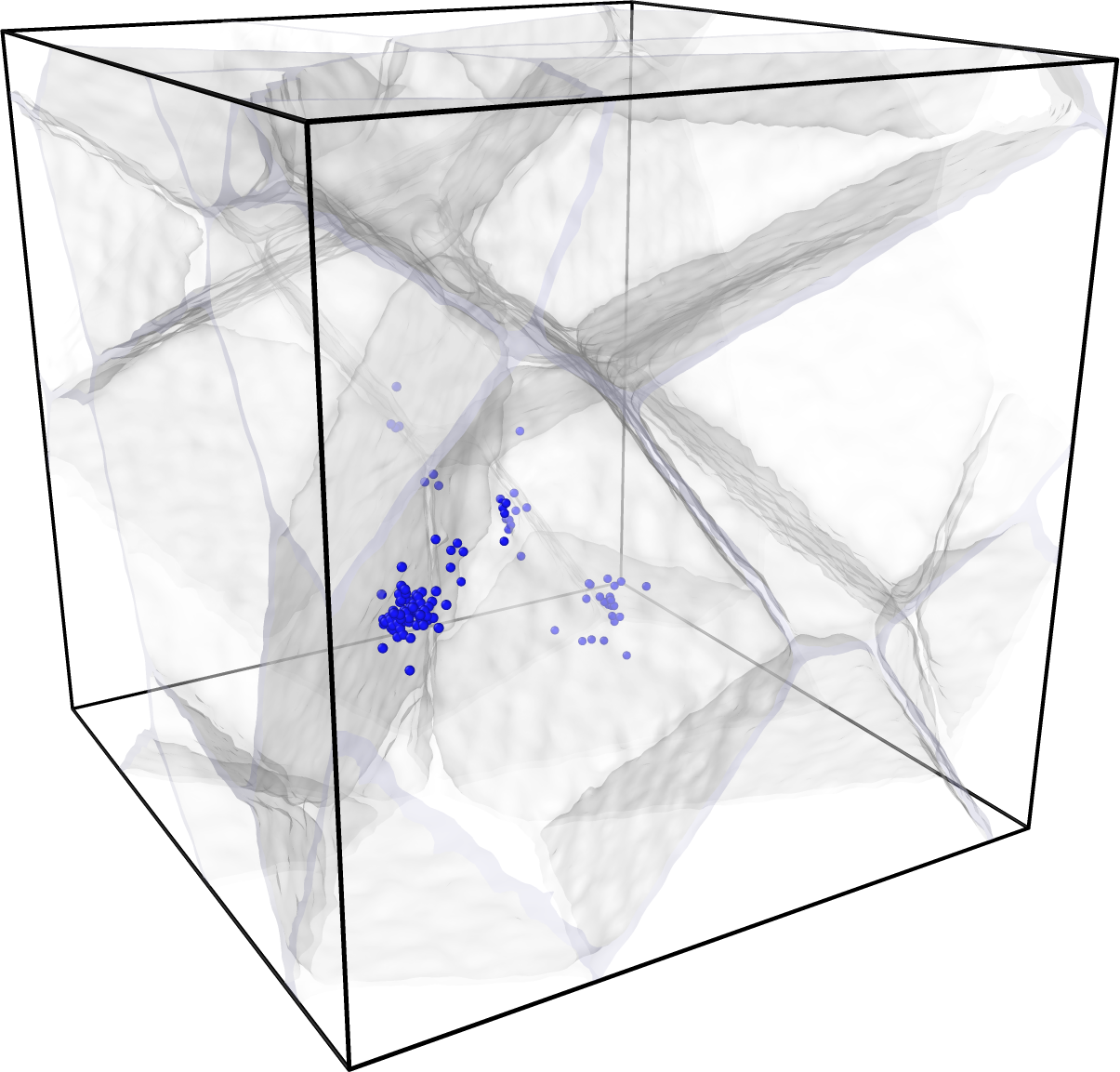}
         \caption{$0$ sec}
         \label{Fig:17a}
     \end{subfigure}
     \begin{subfigure}[b]{0.32\textwidth}
         \centering
         \includegraphics[width=\textwidth]{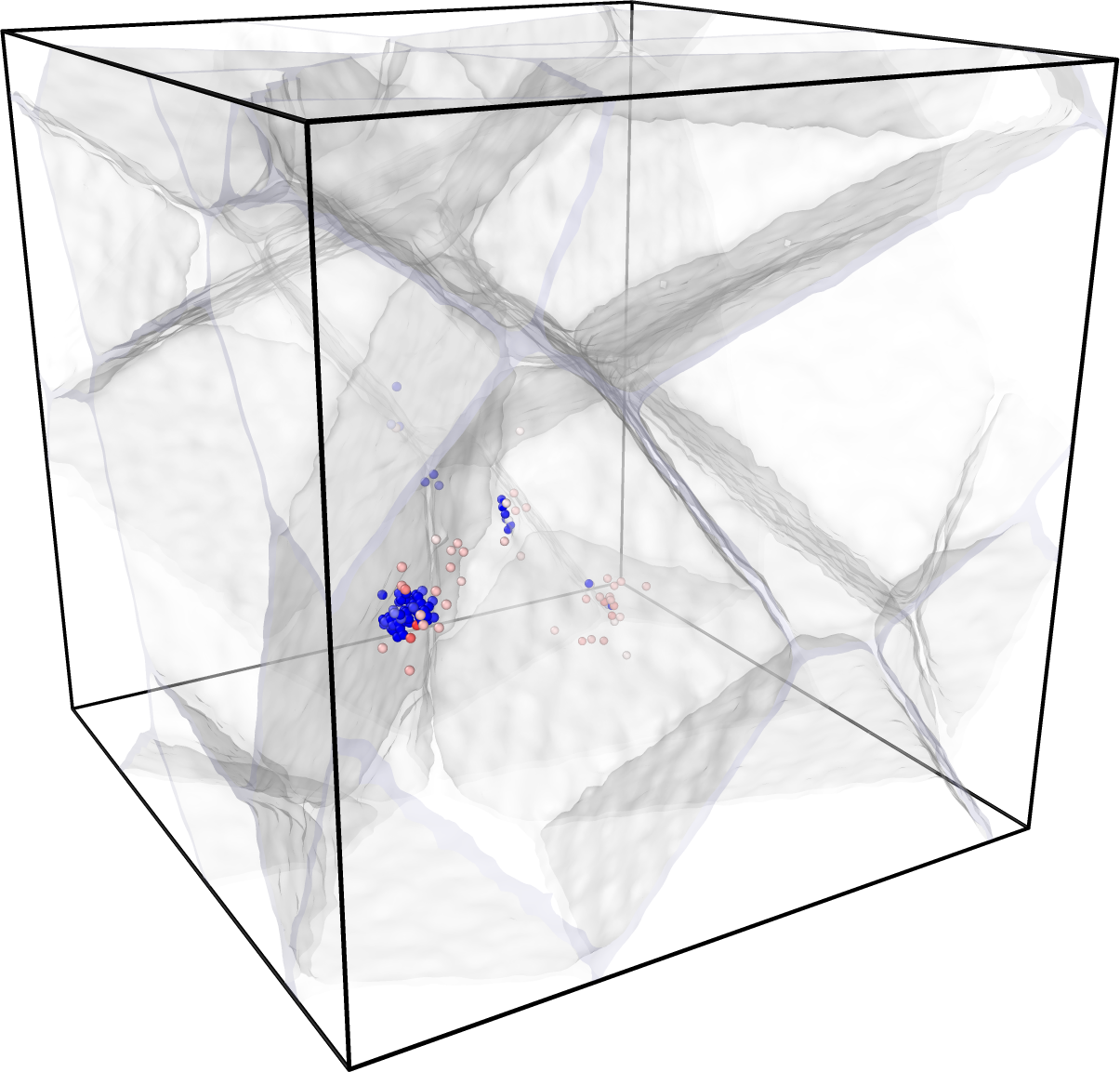}
         \caption{$4.98\times10^{6}$ sec}
         \label{Fig:17d}
     \end{subfigure}
     \begin{subfigure}[b]{0.32\textwidth}
         \centering
         \includegraphics[width=\textwidth]{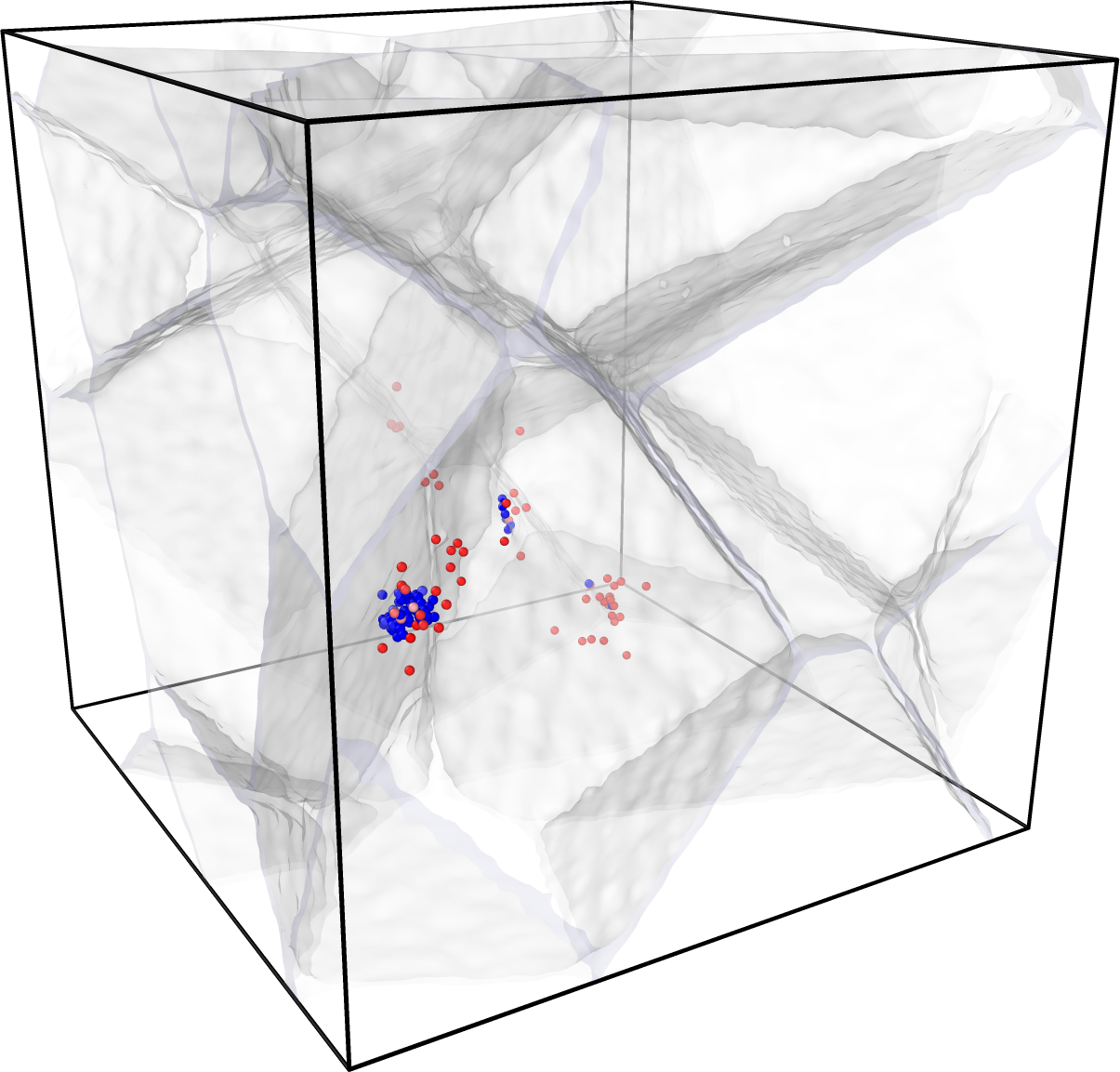}
         \caption{$2.34\times10^{7}$ sec}
         \label{Fig:17f}
     \end{subfigure}
        \caption{Time evolution of the vacancy sites. Vacancies are shown in blue. The gradient of color from blue to red represents the sites change in atomic molar fraction. Sites fully empty ($x_i=0$) are shown in blue. As sites increase their atomic molar fraction they turn to red. Fully occupied sites ($x_i=1$) were removed from the figures. }
        \label{Fig:vac_xi_evolution}
\end{figure}

A spatial representation of this process is illustrated in Fig.~\ref{Fig:vac_xi_evolution} where all FCC atoms were removed from the sample, GB atoms were colored in green, and vacancies with blue for better visualization. For completeness, an animated evolution of the defects analogous to Fig.~\ref{Fig:vac_xi_evolution} is presented in the {\bf Supplementary Video 4}. As time went by, the atomic molar fraction of the vacancies changed with time indicated by the different colors of the original vacancy sites (cf. Fig~\ref{Fig:17a} with Fig~\ref{Fig:17d}). Most of the initial vacancies were removed at the end of the simulation, while another set of atoms remained with low atomic molar fractions, as shown in Fig~\ref{Fig:17f}. Remarkably, most of the surviving vacancies were clustered at several locations of the computational cell.

\section{Discussions} \label{Section:Discussions}

Having presented the results obtained with the $\ell$2T-MD model and the MXE framework, we discuss the main results next. First, let us focus on the results obtained during the PKA simulations with traditional MD and $\ell$2T-MD. From the results presented in Section \ref{Section:PKA} it was evident that traditional MD simulations are not suitable to predict the number, shape, and size of defects induced from the PKA simulations even at low recoil energies, i.e., $\sim50$ keV. This finding is consistent with several other studies where radiation-induced damage has been investigated in SC and NC materials \cite{duffy2006including,Eva:2019,ullah2019new,zarkadoula2019effects,Ullah:2020}. The difference between MD and MD with electronic effects (including SRIM and $\ell$2T-MD) was noticeable in many aspects, including the transient and surviving number of FPs, in the number of defected atoms found in the simulations, and in the dislocation density. The differences in these values increase with the recoil energy and could reach differences of 50\% or more in several metrics (such as dislocation density and number of FPs). The differences between MD and $\ell$2T-MD are due to two effects included in the $\ell$2T-MD model. First, the electronic subsystem can quickly exchange energy with the lattice and possesses heat capacity ($C_e$), allowing it to absorb energy from the PKA very quickly. Second, the electronic subsystem is allowed to dissipate energy by heat conduction, and the thermal conductivity of the electronic subsystem is much faster than the lattice. Thus, the combination of a strong coupling between electrons and phonons (through the electron-phonon constant, $G$ in Eq. \ref{Eq:l2t}), heat capacity ($C_e$), and heat conduction (via $K_{ij}$ in Eq. \ref{Eq:l2t}) results in a much faster energy dissipation than in MD.

Once the electronic effects were assessed in the PKA simulations using single crystals, we investigated the effect of NC samples in the effects of primary radiation damage in Section \ref{Section:PKA-NC}. We found that, independently of the PKA energy, the number of defective atoms {\color{black}around SIAs} and the size of these clusters is always smaller in NC compared to single crystals. {\color{black}While for vacancy clusters, there is no significant difference in terms of number and size of the defects between SC and NC}. Since interstitial atoms have much smaller migration energy than vacancies in bulk, they can quickly move through the NC \cite{bai2010efficient,chen2013defect} to reach the GBs network since the boundaries are only a few nanometers apart. As a result, there are fewer interstitial atoms to recombine than in SC, resulting in smaller SIAs clusters and dislocation loops. An important outcome of the PKA simulations in NC samples was that the number of vacancies was not significantly reduced. Since vacancies have high migration energy ($\sim 0.9$ eV, see Fig. \ref{Fig:16d}), they cannot quickly migrate to the GB network to be annihilated. Thus, only minor differences in the number of vacancies were observed between SC and NC during the PKA simulations.

The use of the MXE simulations allowed us to clarify this point further. In particular, we observed that after long-term behavior, vacancies were able to diffuse in the sample as detailed in Section \ref{Section:MXE}. This long-term diffusion resulted in a drastic reduction of the vacancies (cf. Fig. \ref{Fig:11a} and Fig. \ref{Fig:11b}). {\color{black}An interesting finding was that clusters containing less than $\sim100$ defected atoms annihilated to the GBs with higher percentage compared to the larger clusters. This finding points out that there is a critical cluster size for thermal stability in the long-term behavior of NC samples. While smaller clusters can reduce their size and disappear due to thermal effects, larger ones are stable independently of the simulation time. This behavior agrees with the previous findings that the stability of vacancy clusters strongly varies depending on their size \cite{gilbert2009vacancy, taller2020understanding}. Vacancies diffused to the GB network where they were annihilated as shown in Fig.\ref{Fig:void_evolution}.}

\begin{figure} [H]
     \centering
      \begin{subfigure}[b]{0.46\textwidth}
         \centering
         \includegraphics[width=\textwidth]{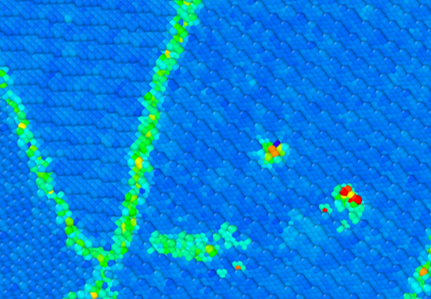}
         \caption{Chemical potential ($\mu_i$) after cascade simulation. }
         \label{Fig:chpot_step0}
     \end{subfigure}
           \begin{subfigure}[b]{0.46\textwidth}
         \centering
         \includegraphics[width=\textwidth]{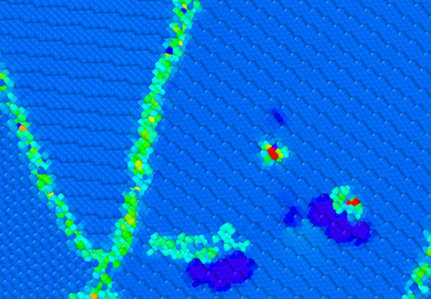}
         \caption{Chemical potential ($\mu_i$) after defects diffusion.}
         \label{Fig:chpot_step30}
     \end{subfigure}
%
     \begin{subfigure}[b]{0.25\textwidth}
         \centering
         \includegraphics[width=\textwidth]{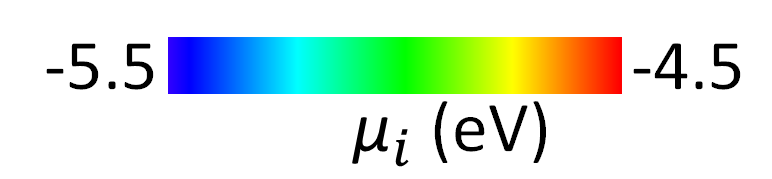}
         \label{Fig:color_map4}
     \end{subfigure}
        \caption{Chemical potential $\mu_i$ for part of the simulation cell after the cascade simulation using a $50$ keV PKA and after the MXE simulation. (a) A slice of the NC simulation cell with the chemical potential of the atoms shown in contour plot. (b) Shows the same slice as in (a) but after defect diffusion with MXE. The color bar at the bottom indicates the chemical potential levels in eV.}
        \label{Fig:chem_pot}
\end{figure}

The diffusion of the defects (vacancies and interstitials) depends on the \emph{local} chemical potential environment (${\bm \mu}_i$), as indicated by the evolution of the Grand canonical free entropy given in Eq. \ref{Eq:Grand-Canonical-Free-Energy} and via Eq. \ref{Eq:MassTransport} for the time evolution of the atomic molar fractions. One of the advantages of the MXE framework is its ability to compute the local chemical potential at each site. To illustrate this feature, Fig.~\ref{Fig:chem_pot} shows the chemical potential for a slice of the NC sample before and after the MXE simulation.
The GB network and the defects originated by the PKA have higher chemical potential relative to the atomic sites in bulk. These variations arise from the local atomic distortion of the NC and long-range elastic effects of point defects (see Fig.~\ref{Fig:chpot_step0}). The chemical potential's gradient is responsible for the diffusion dynamics modeled by Eq. \ref{Eq:MassTransport}.  After the primary PKA damage, we also observed that the chemical potential for atoms in the bulk of the nano-crystals is relatively homogeneous but with some minor fluctuations. 

Figure \ref{Fig:chpot_step30} shows the chemical potential map for the same slice as in Fig.~\ref{Fig:chpot_step0} after the MXE simulation. We first noticed that the chemical potential for all atoms in bulk was homogenous, except near the GBs and surviving defects. However, a few subtle differences appeared after the MXE simulation. For instance, since most defects disappeared or reduced their sizes, the chemical potential showed much sharper gradients and contour. For surviving defects, the shape of the clusters changed to more energetically favorable shaped (more isotropic rather than anisotropic).

Once vacancies reached the GB network, they can quickly diffuse to the most energetically favorable sites since their migration energy is much smaller than the bulk, as shown in Fig. \ref{Fig:16c} and Fig. \ref{Fig:16d}. In the analysis of the vacancy formation and migration energy, we observed that most of the sites in the GB network are more favorable to generate vacancies than the bulk of the NC. At the same time, the vast majority of the sites possess smaller migration energy than the bulk, which ultimately facilitates a much faster diffusion. During the MXE simulation, most of the vacancies in NC are quickly annihilated in the GB network. However, this process reached a saturation point, where several stable vacancies remained in the NC sample. This behavior is attributed to the collective behavior of vacancies and their formation and migration energy. First, most of the surviving defects were clustered of vacancies, as shown in Fig. \ref{Fig:vac_xi_evolution}(c). These clusters of vacancies have much slower diffusivity than individual vacancies, which ultimately slowed down the annihilation of vacancies in the GB network \cite{garcia2002self}. At the same time, while most sites have favorable formation and migration energy, a small percentage of these sites have shown much higher formation and migration energy than the bulk. These unfavorable sites block the diffusion of defects and, in turn, reduce the chemical potential. Since the GB network has many alternative paths, the saturation point is reached slowly, as shown in Fig. \ref{Fig:15}.

The collective diffusive behavior of vacancies can be modeled using the information obtained from MXE. First, let $c=\frac{N_v^t}{N_v^0}$ denote the dimensionless concentration of vacancies in the bulk after the PKA simulation, where $N_v^t$ are the number of vacancies at time $t$ and $N_v^0$ are the number of vacancies at $t=0$ s, and by definition, $c~\in~[0,1]$. The evolution of vacancies can be described by 
\begin{equation} \label{Eq:VacDiff}
\frac{dc}{dt} = - \frac{c}{\tau}
\end{equation} 
where $\tau = R_g \cdot v_d^{-1}$ is a characteristic time for vacancy diffusion to the GB. In the determination of the characteristic time, it was assumed that the shape of the grains is spherical, and therefore, $R_g$ is the grain size radius. $v_d = D_m\cdot b^{-1}$ is a diffusive velocity of vacancies. The radius of the grains can be estimated as $R_g = (\frac{3V}{4\pi N_g})^{1/3}$, where $N_g$ is the number of grain in the cell. The diffusion velocity can be approximate with the use of the vacancy diffusivity, $D_m=\frac{Zb^2}{2d}\nu_0 \exp{\big( -Q_m/k_B T \big)}$. To account for the saturation point of the vacancies, here we recur to the information obtained from MXE. A small portion of these defects remains stable due to their clustering. Thus, the diffusivity of such cluster shall be smaller than of a single vacancy (this could be either due to a reduced migration frequency or higher migration energy of the cluster), since the diffusion of these clusters involves collective motion of atoms, not individual ones. 

To account for this effect, $\tau$ in Eq. \ref{Eq:VacDiff} has to be modified to account for increasing migration energy as the vacancies are annihilated. Thus, the migration energy is changed as a function of the concentration. A trial and error procedure suggested $Q_m(c)= Q_m (1+\gamma/7)$, where $\gamma = (1-c)^2$, where $Q_m$ is specified in Table \ref{Table:material}. The quadratic term penalizes large changes in $c$, allowing individual vacancies to migrate easily at the beginning of the diffusion. As $c$ decreases, $Q_m(c)$ increases rapidly, and so the characteristic time $\tau$, which, in turn, reduces the rate of vacancy annihilation in the bulk. While the model predicts a continuous increment of the migration energy, we expect this increment to be discrete and much sharper than the modeled one. For instance, the migration energy of a double jump increases to 1.5 eV, while the migration energy of a triple jump is around 2.1 eV \cite{garcia2002self}. The comparison between this phenomenological model and the results obtained with MXE are shown in Fig. \ref{Fig:15} with the dashed and dotted lines. In our simulation, $Q_m$ was increased by about 5\%, and we can already see that the system reaches a quasi-steady state. While the model does not exactly match the MXE results, it has all features of the results and the right trends. We also point out that other power laws can modify the migration energy to better match the MXE results. 

The collective spatial and temporal evolution of vacancies have also been investigated, as shown in Fig. \ref{Fig:Xi_evolution_2}. We found that a delta Dirac function can represent the initial distribution of vacancies. However, as diffusion was allowed in the MXE simulation, we found two fascinating behavior of vacancies. First, some sites showed remarkable stability and quickly remained stable during the MXE simulation. These sites represented about $\sim48\%$ of all vacancies generated by the PKA. However, the remaining vacancies sites increased their atomic molar fraction at different rates, and the time evolution of their probability density function was shown in Fig. \ref{Fig:Xi_evolution_2}. Remarkably, when the MXE simulation was stopped, these sites were fully occupied ($x_i=1$), and a delta Dirac function represented their probability distribution function. This behavior resembles similarities with the evolution of probability density functions, mathematically described with the Fokker-Planck equation, similar to the one used to simulate heat and mass transport in this work. 

\section{Conclusions} \label{Section:Conclusions}

We have presented a multiscale and multiphysics framework to simulate radiation-induced damage in nano-crystalline materials. The proposed framework includes electronic effects to simulate the fast interactions between the PKA, electrons, and phonons during radiation events, and the ability to simulate the post-PKA long-term diffusion of defects. These events span from a few fs to hours or days, and they can only be tackle with temporal multiscale frameworks. We found that electronic effects are essential during the PKA events, especially at moderate to high recoil energies. MD simulations that did not incorporate electronic effects overestimated the number and size of defected clusters and dislocation density. SC and NC materials have also been compared. Overall, we found that NC materials can absorb interstitials and vacancies in the GB network, showing a self-healing capability compared with materials with coarser grains. While both interstitials and vacancies are absorbed in the GB network, the time scale for these events is highly dissimilar. Most interstitial atoms were reabsorbed during the primary radiation damage due to the smaller formation and migration energies. 
Overall, the simulations performed with the MXE allowed us to elucidate the self-healing behavior of NC samples. We also found that this effect is, however, limited in the long term. Since a small percentage of vacancies were clustered after the PKA, the migration energy of these clusters is much higher than individual vacancies. In addition, a small percentage of atoms in the GB possessed higher vacancy formation and migration energies, slowing down diffusion until a saturation point was reached. About $50\%$ of all vacancies generated during the PKA remained stable in the NC sample.

The proposed multiscale framework can be used to investigate the effects of the grain size, geometry, and chemical composition in the radiation-induced damage in other nano-crystalline materials. For instance, multi-compositional element alloys, including medium and high-entropy alloys, could be investigated with the proposed framework. The authors actively pursue this research line.

\section{Acknowledgements} 
We gratefully acknowledge the support from the Natural Sciences and Engineering Research Council of Canada (NSERC) through the Discovery Grant under Award Application Number 2016-06114, the New Frontiers in Research Fund (NFRFE-2019-01095), and the computational resources provided by Compute Canada and the Advanced Research Computing (ARC) at the University of British Columbia. M. H. gratefully acknowledges the Four Year Fellowship program granted by the Department of Mechanical Engineering at the University of British Columbia.

\bibliographystyle{unsrt}
\bibliography{ref.bib}

\end{document}